\documentclass[]{mn2e} 
\usepackage{graphicx}	
\usepackage{bm}  
\usepackage[T1]{fontenc} 
\DeclareRobustCommand{\VAN}[3]{#2}
\let\VANthebibliography\thebibliography
\def\thebibliography{\DeclareRobustCommand{\VAN}[3]{##3}\VANthebibliography}
\usepackage{graphicx}	
\usepackage{amsmath}	
\usepackage{amssymb}	
\title[Origin of the Galactic-Centre Filaments]{Supernova-remnant origin of the Galactic-Centre filaments
}

\author[Yoshiaki \textsc{Sofue}]{Yoshiaki \textsc{Sofue}\\
Institute of Astronomy, The University of Tokyo, Mitaka, Tokyo 181-0015}

\def\vrot{V_{\rm rot}}   

\def\Msun{M_\odot} 
\def\deg{^\circ} 
   
\def\be{\begin{equation}} \def\ee{\end{equation}}

\def\kms{km s$^{-1}$}
 \def\epc{{\rm ~pc~}}
 \def\rmh2{{\rm H_2}}

\def\Alf{Alfv\'en } \def\alf{Alfv\'en }

 \def\Vu{V_{\rm unit}} \def\vu{V_{\rm unit}}  \def\Bu{B_{\rm unit}} \def\rhou{\rho_{\rm unit}} 
 \def\tu{t_{\rm unit}}\def\lu{L_{\rm unit}}
 \def\Va{V} \def\va{V}
\def\pr{p_r}
\def\d{\partial}
\def\pth{p_\theta}
\def\pphi{p_\phi}
\def\cot{{\rm cot}\ }
\def\sin{{\rm sin}\ }

\def\f{\frac}
\def\apjl{ApJ. L.}\def\aap{A\&Ap}\def\aaps{A\&Ap. Suppl.}\def\mnras{MNRAS}\def\pasj{PASJ}\def\apj{ApJ}
\def\epc{{\rm pc}}
\def\rad{R}

\begin{document} 
\maketitle   

\begin{abstract}   
The mechanism to produce the numerous Galactic-Centre filaments (GCF) that vertically penetrate the Galactic plane {without clear evidence of connection to the disc} remains a mystery.
Here we show that the GCFs are explained by relics of supernova remnants (rSNR) {driven by hundreds of supernovae (SNe)} exploded in the star-forming ring of the central molecular zone (CMZ) at an SN rate of $\sim 2\times 10^{-4}$ y$^{-1}$ in the past $\sim 0.5$ My.
The evolution of rSNRs is simulated by the propagation of fast-mode magneto-hydrodynamic (MHD) waves, which are shown to converge around the Galactic rotation axis by the focusing effect.
Tangential projection of the cylindrical wave fronts on the sky constitutes the vertical filaments.
The SNR model explains not only the morphology, but also the nonthermal radio spectrum, smoothed brightness over the distribution area consistent with the $\Sigma-D$ relation of SNR, and the heating mechanism of hot plasma in the GC.
We also discuss the implication of this model on the study of the interstellar physics and star-formation activity in the Galactic Centre.
\end{abstract} 
 

\section{Introduction}

The high-resolution radio continuum observations with the VLA in the decades
 \cite{yu+1984,mo+1985,an+1991,la+1999a,laro+2004,pa+2019}
and more recently by MeerKAT
(Heywood et al. 2019, 2022; Yusef-Zadeh et al. 2022a, 2022b)
have revealed the numerous Galactic-Centre filaments (GCF) that vertically penetrate the Galactic plane without {clear evidence of interaction with the disc}.
The forest of GCFs runs predominantly inside the GC lobe (GCL) \cite{so+1984,hey+2019}, drawing a large "harp" shape from $l\sim -1\deg$ to $+0\deg.3$ and $b\sim -1\deg$ to $+1\deg$.
The major filaments are often bifurcated into thinner strings {\cite{hey+2022,yu+2022b}}, composing a number of "mini harps".
Individual filaments exhibit more complicated morphology, curving, crossing, horizontal{ \cite{la+1999a,yu+2022a}}, and partial loops.
There are also many filaments running outside the GCL, covering a wide area from $l\sim -1\deg.5$ to $+1\deg$ ($\sim -200$ to $\sim +150$ pc).
The GCF is also called non-thermal filament (NTF), non-thermal radio filament (NRF), thread, or string (the literature as above).

There have been several models to explain the origin of the GCF as due to outflow and ejection from the magnetized central disc \cite{da+2002,yu+2019,ba+2019,zhang+2021,yu+2022c}, or by interaction with high energy objects such as pulsar winds \cite{bo+2006}.
They attribute the driving momentum and energy to outflow activities in the disc or to direct energy injection from the sources.
These mechanisms require that the roots of GCF are connected to the galactic disc or to high-energy driving sources. 
However, the majority of the filaments appear to have no clear evidence of association with such driving objects, although some possible correlations of filaments with compact radio sources are suggested \cite{yu+2022c}.
Therefore, the mechanism that produces the apparent free penetration through the disc and the unique 'harp string'-like morphology seems to be still shrouded in mystery.

In order to solve this problem we have proposed a model which attributed the GCF to magneto-hydrodynamic (MHD) disturbances excited by the central activity near Sgr A in a vertical magnetic cylinder{ \cite{so2020th}}, which was partially successful to explain the GCF morphology.
In this paper we extend this idea, and propose a more plausible mechanism to produce the GCF by supernovae (SN) occurred in the star-forming (SF) ring of the central molecular zone (CMZ). {The GCFs are here interpreted as 'relic' of supernova remnants (rSNR) and are approximated by MHD disturbances of small amplitude propagating at the \alf velocity.  }
 
\section{Propagation of MHD Waves}  

\subsection{Fast-mode MHD waves}

MHD disturbances excited by explosive events as SNe in the interstellar space propagate as spherical shock waves in the initial expansion phase. In the fully expanded phase, they propagate as the fast-mode MHD compression wave (hereafter, MHD wave), \alf wave, and the sound wave. Among the three modes, the fast MHD wave is most efficient to convey the energy released by the rSNR into the ISM in the present circumstances \cite{so2020th}, which propagates at the group velocity equal to the \alf velocity, $v_{\rm g}\simeq \va$ \cite{porter+1994,kumar+2006}.
{Predominance of MHD wave is confirmed by the low value of plasma $\beta=(c_{\rm s}/\va)^2\sim 0.01$, as $\va\sim 10^3$ \kms and $c_{\rm s}\sim 10^2$ \kms in the present circumstance.}

\begin{figure}   
	\begin{center}
	\includegraphics[width=7cm ]{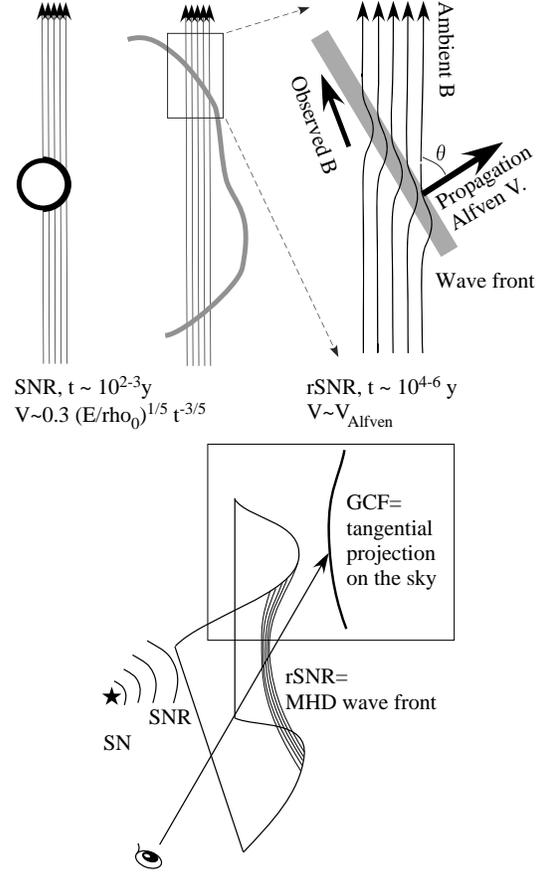}   
\end{center}
\vskip -3mm
\caption{{Schematic illustration of rSNR as a fast-mode MHD wave front. An SN produces an SNR expanding as a Sedov sphere. As it fully evolves, the shell is strongly deformed and merged into the ISM as an MHD wave (disturbance) propagating at \alf velocity (sub-Alfv\'enic). The tangential projection of the front on the sky is observed as a GCF.  } }
\label{illustrSNR} 
	\end{figure}

{Here, we assume that an rSNR after sufficient expansion of an SNR, which should actually be a shock wave, is represented by an MHD wave propagating at the \alf velocity.
So, rSNR is no more a shocked shell, but is sub-Alfv\'enic wave.
Fig. \ref{illustrSNR} schematically illustrates the  assumed situation in this paper.
The wavelength is assumed to be comparable to the shock thickness, which is on the order of $\lambda\sim 0.1-0.3 r_{\rm shock}$ for a usual SNR or adiabatic shock-wave shell of radius $r_{\rm shock}$. 
We assume that the amplitude of the wave is infinitesimal, so that the waves are treated by WKB (Wentzel-Kramers-Brillouin) approximation \cite{uch1974}. By this approximation, however, we cannot compute the nonlinear compression of the gas and magnetic field, and therefore radio emission. 
So, the present work is mainly aimed at morphological studies of the shape of GCF.} 
	
The propagation of MHD waves is traced by solving the Eikonal equations developed for the Solar coronal Moreton waves{ \cite{uch1970,uch1974}}.
The method has been applied to the GC and SNRs  \cite{so1977,so1978,so1980,so2020th,so2020fe}.
{In the interstellar space, a gas cloud with higher density and lower \alf velocity acts as a convex lens (low-$\va$ lens) gathering the waves, while a gaseous hole with higher magnetic strength acts as a concave lens (high-$\va$ lens) reflecting the waves \cite{so1978}. As discussed later, the hot plasma in the GC filling the vertical magnetic cylinder (low-$\va$ pillar) acts as a convex toric lens gathering the waves toward the cylinder axis. }

The Eikonal equations describing the propagation of an MHD wave packet are given as follows \cite{uch1970,uch1974}.
\be \f{dr}{dt}=V \f{\pr}{p}, \ee
\be \f{d\theta}{dt}=V \f{\pth}{rp}, \ee 
\be \f{d\phi}{dt}=V \f{\pphi}{rp\  \sin \theta}+\Omega, \ee 
\be \f{d\pr}{dt}=-p\f{\d V}{\d r} +\f{V}{rp}(\pth^2+\pphi^2), \ee
\be \f{d\pth}{dt}=-\f{p}{r} \f{\d V}{\d \theta} 
- \f{V}{rp}(\pth \pr-\pphi^2 \ \cot \ \theta), \ee 
\be \f{d\pphi}{dt}=-\f{p}{\sin \theta} \f{\d V}{\d \phi} - \f{V}{rp}(\pphi \pr + \pphi \pth \ \cot \ \theta), \ee 
where
$V=V(r,\theta,\phi)=V(x,y,z)$ is the \Alf velocity, the vector $\bm{p}=(\pr, \pth, \pphi)={\rm grad}\  \Phi$ is defined by the gradient of the eikonal $\Phi$, $p=(\pr^2+\pphi^2+\pth^2)^{1/2}$, $(r,\theta, \phi)$ and $(x,y,z)$ are the polar and Cartesian coordinates, and $\Omega$ is the angular velocity of the ambient material around the Galactic rotation axis ($z$ axis).   
 
\subsection{Magnetic view of the GC and \alf velocity}

\begin{figure}
	\includegraphics[width=7cm ]{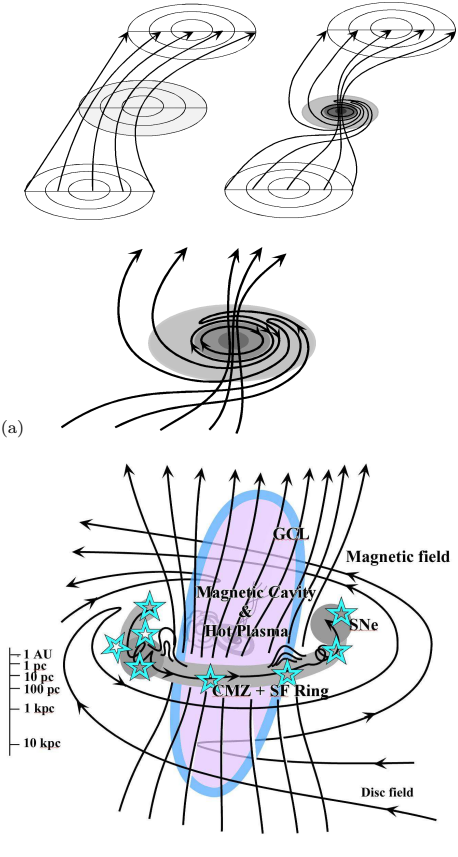}  
\begin{center} 
\end{center}
\vskip -6mm
\caption{Magnetic view of the Galactic Center (Sofue et al. 2010).
(a) Formation of the ring, spiral and vertical magentic fields in a primordial galaxy moving through the intergalactic magnetic field. 
(b) Magnetic view of the GC with the cavity, hot plasma, CMZ, star-forming (SF) ring, and SN explosions. } 
\label{magview} 
	\end{figure}

We assume an oblique dipole-like magnetic cylinder \cite{so+2010,so2020th} filled with a hot plasma{ \cite{nax+2019,po+2019,po+2021}}, which are co-existing with the central molecular zone (CMZ) and star-forming (SF) ring in the GC \cite{hen+2022,so2022}.
Fig. \ref{magview} schematically illustrates the assumed magnetic configuration and gas density distribution based on the primordial magnetic field scenario \cite{so+2010}.

The \alf velocity is calculated from the magnetic field $B$ and gas density $\rho$ by
\be V=\sqrt{B^2/4\pi \rho}. \ee
The gas density is expressed by superposition of a molecular ring \cite{so1995,hen+2022}, hot plasma filling the magnetic cylinder \cite{nax+2019,po+2019}, and an extended halo with constant density as follows.
\be \rho=\rho_{\rm ring}+\rho_{\rm hot}+\rho_{\rm halo}, \ee
where
\be \rho_{\rm ring}=\rho_1\exp[-((\varpi-r_{\rm ring})/w_{\rm ring})^2+(z/w_{\rm ring})^2], \ee 
\be \rho_{\rm hot}=\rho_2 \exp(-\xi^4), \ee
and 
\be \rho_{\rm halo}=\rho_3. \ee 
Here, {$(x,y,z)$ are cartesian coordinates with the origin at the GC and $x$ denotes distance projected on the sky toward the east (positive longitude), $y$ toward the anti-center direction, and $z$ is the polar axis (rotation axis) toward the northern Galactic pole. 
The radius 
\be\varpi=\sqrt{(x-q z)^2+y^2}\ee 
represents distance from the tilted cylinder axis inclined by $q={\rm tan}\chi \sim 0.2$ from the $z$ axis with $\chi\sim 10\deg$, {and $\xi$ represents the shape of the elliptical cavity,}
\be \xi=\sqrt{\varpi/r_{\rm cav})^2+(z/z_{\rm cav}))^2}.\ee
We assume that $\rho_1=10\rhou$ for the molecular ring, 
$\rho_2=0.1\rhou$ for hot plasma in the magnetic cavity, 
$\rho_3=0.01\rhou$ for uniform halo,
$r_{\rm cav}=1\lu$ and $z_{\rm cav}=3\lu$ for the hot cavity's radial and vertical radii with the density and length units described below.}

	\begin{figure}  
\begin{center}     
(a)\includegraphics[width=7cm ]{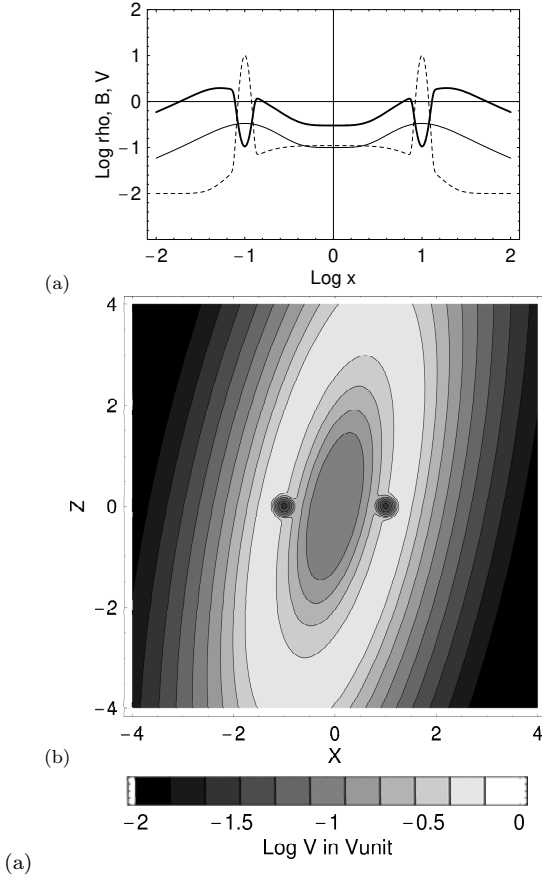} 
\end{center} 
\caption{
(a) Model profiles of the gas density log $\rho$ (in $\rhou=$ 1 H cm$^{-3}$: dashed line), magnetic field strength log $B$ (in $\Bu=1$ mG: thin line), and \alf velocity log V (in $\Vu=2183$ \kms: thick line), and $x$ is in $\lu=100$ pc.
(b) Model distribution of the \alf velocity in $\Bu$ in the $(x,z)$ plane. 
} 
\label{magrho} 
	\end{figure}

The magnetic strength is assumed to have an oblique dipole-like distribution with a cavity filled by the hot plasma elongated in the polar direction, which is expressed as
\be B =B_0 D(\varpi,z)C(\varpi,z) \ee
where
\be D(\varpi,z)=1/(1+\zeta^4) , \ee
represents the dipole-like magnetic strength and
\be C(\varpi,z)=1-\eta \exp(-\zeta^4) \ee 
represents the central cavity, where $\eta\sim 0.9$ and
\be \zeta=\sqrt{\varpi/r_{\rm mag})^2+(z/z_{\rm mag}))^2}. \ee
Fig. \ref{magrho} shows the longitudinal variations of the assumed gas density, magnetic strength, and Alfven velocity in the Galactic plane.

We assume $r_{\rm ring}\sim 1\lu$ and $w_{\rm ring}\sim 0.05-0.1 \lu$ for the ring, and 
$r_{\rm cav}=r_{\rm mag}\sim 1\lu$ and $z_{\rm cav}=z_{\rm mag}\sim 3\lu$ 
for the magnetic cylinder with cavity and the filling hot plasma.
The \alf speed has the maximum near the molecular ring and a broad minimum in the center (Fig. \ref{magrho}).
Due to the high density in the molecular ring, $\va$ has also a deep minimum in the ring.   

\subsection{Units}

The real quantities are obtained using the units of length $\lu$, time $\tu$, and velocity $\vu=\Bu/\sqrt{4\pi \rhou}$.
Here, the following units are assumed: 
$\rhou=1$ H cm$^{-3}$, $\Bu=1$ mG, $\lu=100$ pc, leading to 
$\Vu=2183$ \kms and $\tu=1\lu/\Vu=4.482\times 10^4$ y.

For a typical magnetic strength in the GC of $\sim 0.1$ mG \cite{yu+2022a}, the \Alf velocity is $\Va \sim 72$ \kms in the disc (CMZ) with $\rho\sim 10 \rhou$, and $\va\sim 690$ \kms in the hot plasma with $\sim 0.1 \rhou=0.1 $ H cm$^{-3}$.
The time scale for a wave travel from the SF ring to the GC through  hot plasma is 
$t_{\rm scale}\sim 100 {\rm pc}/690 {\rm km~ s^{-3}}\sim 1.4 \times 10^5$ y.  

{The \alf velocity may be compared with the sound velocity in the CMZ with $c_{\rm s}\sim 10$ \kms for $T\sim 10^4$ K, and in the hot plasma with $c_{\rm s}\sim 10^2$ \kms for $T\sim 10^6$ K.
Hence, the plasma $\beta$ value is as low as $\beta \sim 0.02$ both in the CMZ and hot plasma, which confirms that the fast-mode MHD wave approximation is reasonable.} 

\subsection{{Dissipation of the wave}} \label{secdissipation}

The fast-mode MHD wave of compression mode propagates across the magnetic field lines at the group velocity approximately equal to the \Alf velocity.
The wave suffers from the viscous and Ohmic energy losses, while the latter is much smaller and negligible, so that the wave is dissipated at the rate $\nu_{\rm vis}$ {\cite{porter+1994,kumar+2006}} approximately given by
\be
\frac{\nu_{\rm vis}}{\rm s^{-1}}
\simeq 10^{7.3} \left(\frac{T}{K}\right)^{5/2} 
\left(\frac{\rho}{\rm H\ cm^{-3}}\right)^{-1} 
\left(\frac{k}{\rm cm^{-1}}\right)^2 {\rm sin}^2 \theta.
\ee
Here, $k=2\pi/\lambda$ is the wave number, and $\lambda$ is the wavelength. 
The damping length (distance) is defined through the group velocity $v_{\rm g}$ and $\nu_{\rm vis}$ by
$L_{\rm damp}=2v_{\rm g}/\nu_{\rm vis}\simeq 2\va/\nu_{\rm vis}$ \cite{porter+1994}.
We here rewrite it in terms of $B$, $\rho$ and $T$ as a function of $\lambda$.
\be
\frac{L_{\rm damp}}{\epc} 
\sim 1.71\times 10^3 
\nonumber
\ee
\be
\times
\left(\frac{T}{10^6 {\rm K}}\right)^{-5/2}
\left(\frac{B}{\rm mG} \right)
\left(\frac{\rho}{\rm H\ cm^{-3}}\right)^{1/2}
\left(\frac{\lambda}{\epc}\right)^2  
{\rm sin}^{-2} \theta.
\ee 

\begin{figure}  
\begin{center}     
\includegraphics[width=7.5cm ]{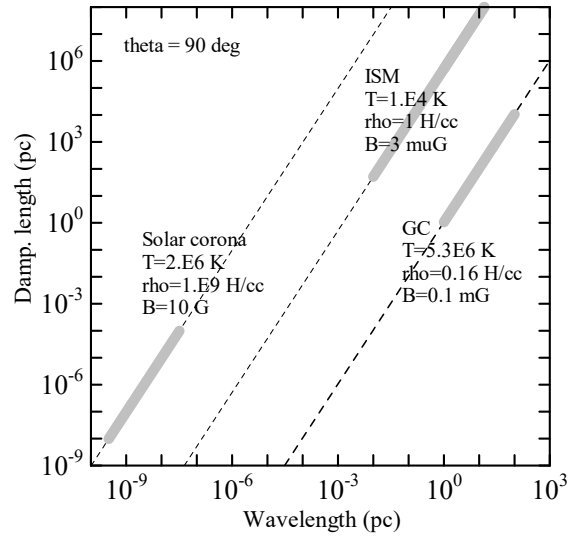} 
\vskip -3mm
\end{center} 
\caption{Damping length $L_{\rm damp}=2 V/\nu_{\rm vis}$ following Porter et al. (1994) as a function of the wave lengths for $\theta=90\deg$. 
The lower line shows that in the GC for 
$T=5.3\times 10^6$ K, 
$\rho=0.16$ H cm$^{-3}$, and
B=0.1 mG ($V=545$ \kms).
The middle and upper dashed lines show those for the interstellar medium with $T=10^4$ K, $\rho=1$ H cm$^{-3}$ and $B=3~\mu$G, and for the solar corona with $2\times 10^6$ K, $10^9$ H cm$^{-3}$, and 10 G ($V=1000$ \kms; Porter et al. 1994), respectively.
The grey thick lines represent possible ranges for the rSNR in the GC, ISM and solar corona.  } 
\label{damping} 
	\end{figure}

In Fig. \ref{damping} we plot the calculated damping length against wavelength for $\theta=90\deg$, where
we adopted the plasma density and temperature from the Suzaku X-ray observations \cite{nax+2019},
$\rho=0.16$ H cm$^{-3}$ and 
$T=0.46 \ {\rm keV}=5.3\times 10^6$ K, and assume $B=0.1$ mG ($\va=545$ \kms).
The plot may be compared with that for the solar corona assuming $T=2\times 10^6$ K, $\rho=10^9$ H cm$^{-3}$ and $B=10$ G ($\va=1000$ \kms) \cite{porter+1994}.
We also show the case for the interstellar medium with $10^4$ K, 1 H cm$^{-3}$ and $3~\mu$G.
The grey thick lines represent possible ranges of the wavelength representing the scale length of the disturbances, and the corresponding damping distances. 
The wavelength is here taken as the representative size of the actual shock thickness of the rSNRs after they have fully expanded in the central $\sim 100$ pc region, where it is on the order of $\lambda\sim 0.1 - 0.3r_{\rm shock} \sim 3$ to 30 pc.
The plot indicates that the damping length is $L_{\rm damp}\sim 100$ pc for $\lambda\sim 10$ pc and greater for longer $\lambda$. The length will be even longer when the propagation is oblique ($\theta\gtrsim 90\deg$) due to the term ${\rm sin}^{-2}\theta$. 

\section{SNR model for the GCF}

\subsection{Vertical filament as tangential projection of wave front}

\begin{figure}  
	\begin{center} 
\includegraphics[width=6.5cm]{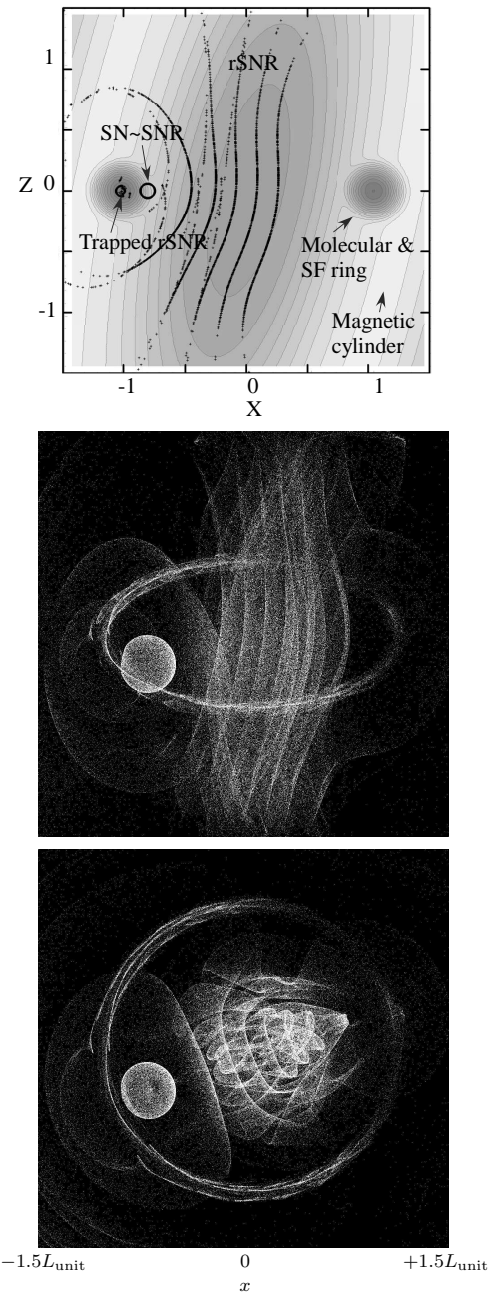}  
\end{center}  
\caption{{[Top] Illustration of evolution of an  rSNR near the CMZ and magnetic cavity filled by hot plasma.
[Middle] MHD wave fronts from one SN exploded at the center of the sphere around $x=0.9\lu$ which represents the front at $t=0.1\tu$.
The density of dots projected on the sky simulates the brightness of an rSNR. 
Each dot represents one wave packet, and the wave front is represented by $10^4$ packets. 
The front is shown at $t=0.1,\ 1,\ 2,\ 3, ... 10\tu$, and projected on a plane 30 degrees tilted from the $z$ axis.  
[Bottom] Same, but projection on the $(x,y)$ plane (top view). The focusing effect is evident toward the rotation axis with the concave fronts converging to the focal point in the opposite side of the axis.} }
\label{805r1} 
	\end{figure}

Fig. \ref{805r1} shows the result of computation for a spherical MHD wave produced at a point near the molecular ring in a non-rotating medium.
{The top panel illustrates the relation of the propagation of the front represented by the strings of dots and the background $\va$ distribution in the $(x,z)$ plane, where an SN exploded at $(x,y,z)=(-0.9,0,0)\lu$.}
In the 2nd and 3rd panels, the wave front is expressed by $1 \times 10^4$ wave packets (later occasionally 1 to $4 \times 10^4$ packets per one rSNR, where the SN expleded at $(x,y,z)\sim (-0.8,0,+0.1)\lu$.
Each of the $1-4\times 10^4$ wave packets is given random radial direction at the explosion center and propagate radially making an expanding spherical front, representing a point (SN) explosion. 
Soon after the size becomes comparable to the ring and wall sizes, it is rapidly stretched in the $z$ direction along the vertical magnetic wall.
Then, the front is refracted and converged toward the rotation axis ($z$-axis) by the focusing effect due to the decreasing \alf speed.
The curved front projected on the sky composes a vertical sharp edge, mimicking the radio filaments.
After passing the cylinder, the front is reflected by the magnetic wall on the opposite side, returns to its coming direction, and finally dilutes and looses the shape.
Besides the expanding and deforming wave, some fraction of the front is trapped by the molecular ring, and makes a belt of filaments which are stretched, tangled, and spiralling along the star-forming (SF) and molecular gas ring.

The figure simulates the propagation of a single MHD front from the explosion origin.
However, it also represents a case when multiple explosions take place periodically at the same place, or a case that a cluster of SNe happens in a small region.
The tangential projection of the fronts appear as multiple parallel filaments, whose separation is proportional to the time interval of SN explosions.
As discussed later, a cluster of SNe in a small area around an star-forming (SF) region at short time interval and limited duration, or a mini starburst as taken place in Sgr B2 \cite{ha+1994}, produces a bunch of short-separation filaments, or a "mini harp".

\subsection{MHD wave front model and properties of simulated filaments}
	
\begin{figure} 
\begin{center}     
\includegraphics[width=7.5cm]{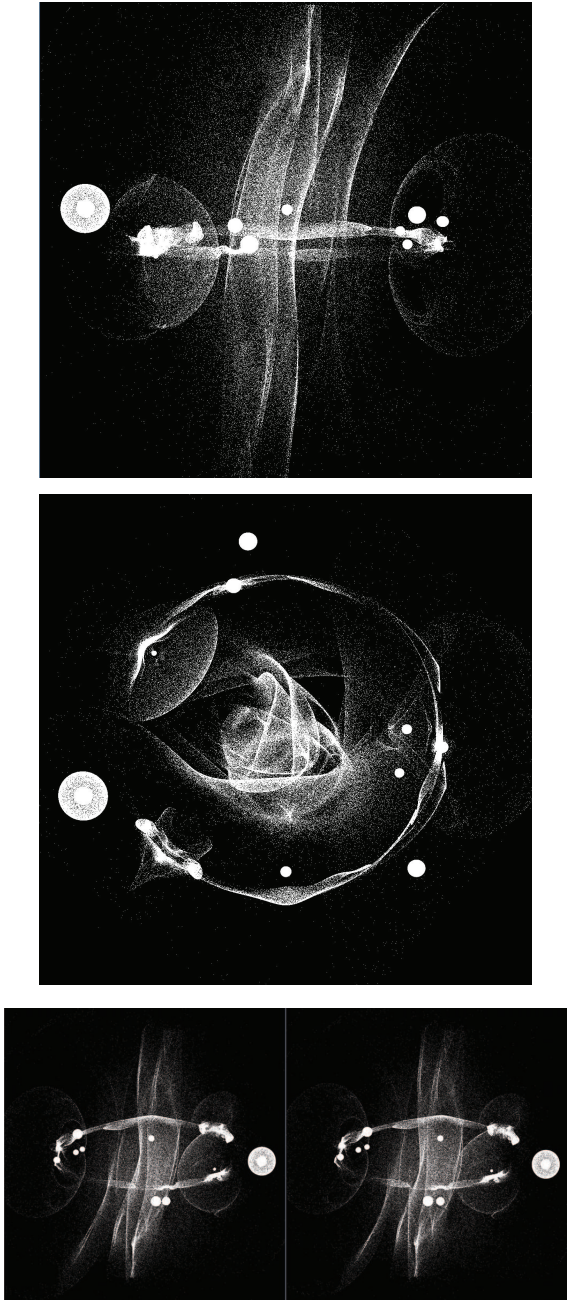}   
\end{center}
\caption{ 
[Top] Side view seen from $5\deg$ above the $(x,y)$ (galactic) plane of the simulated MHD wave fronts excited by 10 SNe exploded randomly near the SF ring. The shells denotes the fronts at $t=0.1\tu$, and the extended fronts are those of 10 rSNRs at $t=10\tu$.
The box area size is $3\lu\times 3\lu$.
[Middle] Same, but a top view.
[Bottom] Same, but from different azimuth angle for stereo-graphic view.
} 
\label{813} 
	\end{figure}

We then simulate a more realistic case on the assumption that SF activity in the molecular (SF) ring resulted in several SN explosions. 
Fig. \ref{813} shows an example of the simulation for ten SNe explosions several $\tu$ ago at random near the SF ring of radius $r=1\lu$, where the interstellar medium is not rotating.
The initial positions of the explosions are marked by the thick shells whose sizes indicate the density of the surrounding medium with larger shells denoting faster \alf velocity.
The bottom panel shows two images seen from 5-degrees different azimuth directions for a stereo-graphic view of the wave fronts.

\begin{figure} 
\begin{center}     
\includegraphics[width=7cm]{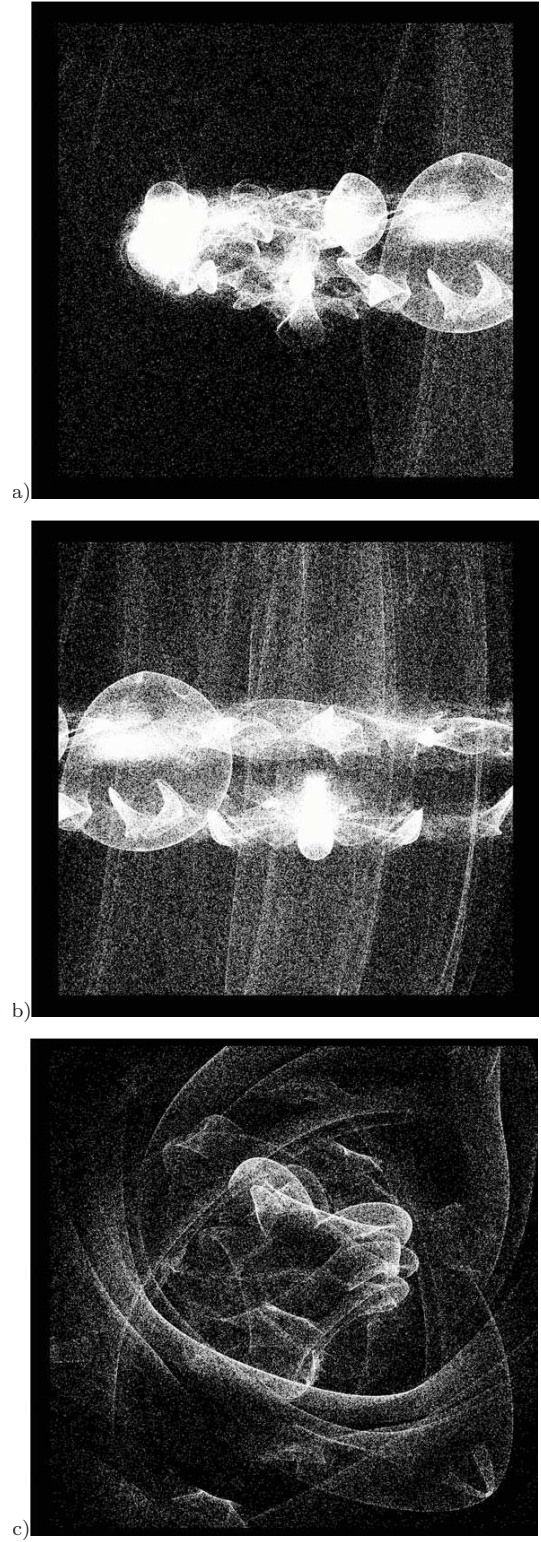}  
\end{center}
\caption{[Top] Close up of the MHD waves near the ring edge, [middle] central region, and [bottom] face-on view of the center.  The area size is $1\lu\times 1\lu$. } 
\label{closeup} 
	\end{figure}

Fig. \ref{closeup} shows a close up of the simulated MHD fronts, where the SF ring's width is taken wider of $0.2\lu$.
Some fraction of the waves are trapped by the molecular ring due to the low \alf velocity, and propagate inside the ring, composing complicated front shapes. They may further affect the SF activity and cloud formation.

\begin{figure} 
\begin{center}     
\includegraphics[width=7.2cm]{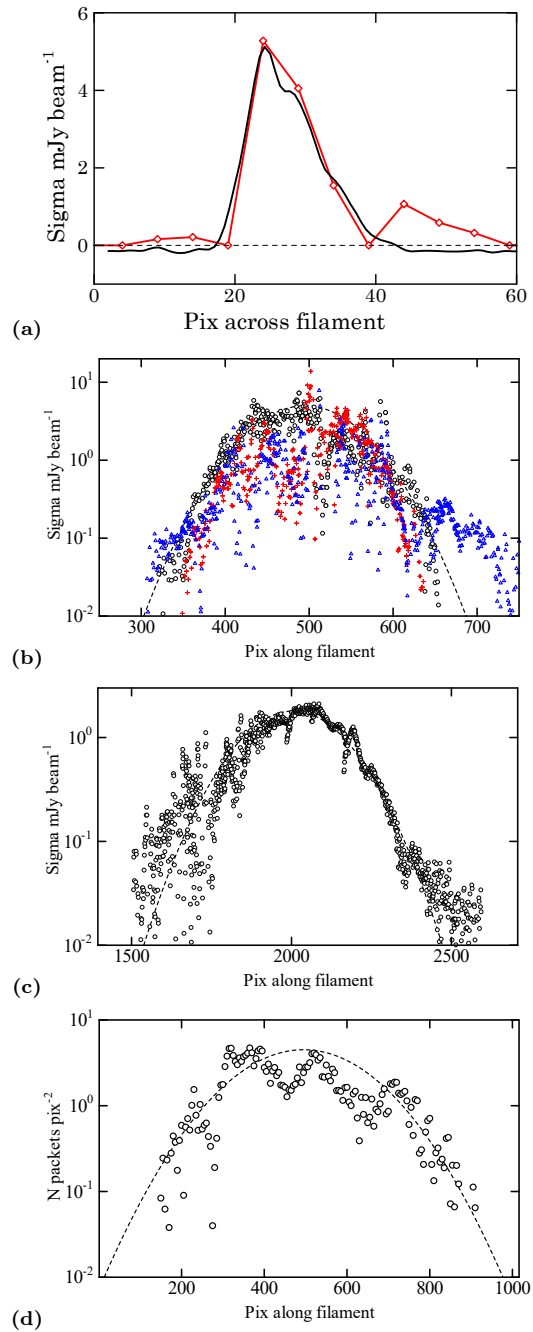}  
\end{center}
\caption{(a) Cross section of the filament G-0.25+0.20 in the MeerKAT 1.28 GHz BGF map (black line) and of a MHD front (red). MeerKAT intensity scale is radio surface brightness and axis scale is relative in pixels. The simulation's intensity scale is proportional to the number density of wave packets adjusted to the radio intensity. 
(b) Radio intensity distribution along the three brightest filaments in the Radio Arc (circles: the brightest near the arc center; crosses: 2nd brightest: triangles: 3rd and longest). Axis scale is relative in pixels.
(c) Same, but the longest filament in the MeerKAT field across G0.10+0.10.
(d) Same, but a simulated filament for by the rSNR model. 
Dashed lines in (b) to (c) are approximate fits by Gaussian functions.
Note the similarity between the observed and simulated distributions.]
} 
\label{filament} 
\end{figure}

In Fig. \ref{filament} (a) we show a cross section of the radio brightness at 1.3 GHz for an isolated filament at G$-2.5+2.0$, and compare with the cross section of one of the simulated filaments.
The simulation well reproduces the sharp rise from the left side (convex side) followed by a gradual decrease toward the right.
Such a shape is typical for sky projection of a curved thin surface.
The surface is either expanding from the right (west) to left (east), or converging (focusing) from the left to right.  

  We also compare the radio intensity distributions along the GCF with the simulation.
Fig. \ref{filament} (b) shows the three brightest filaments of the Radio Arc, and panel (c) the longest filament in the MeerKAT GC field running across G0.10+0.10. 
The radio intensity mildly increases from one end, attains a broad maximum near the central part, and again mildly decreases toward the other end.
The intensity variation near the maximum is mainly due to fore- and background thermal filaments.
The shape is roughly symmetric with respect to the maximum, and is approximately fitted by a Gaussian function as indicated by the dashed lines.
It may be noticed that there appears no particular exciting source on the filaments.
In panel (d) we show the simulated intensity distribution along a typical isolated filament in the MHD wave front based on the SNR model, which shows a similar profile to the observations. 
The variable intensity near the maximum as observed in panels (b) and (c) is also simulated as due to overlapping filaments and front corrugation.
The similarity of the intensity distributions between the observation and simulation both across and along the filaments would be in favor of the present sheet model. 
	
\subsection{Effect of the Galactic rotation}

\begin{figure*} 
\begin{center}     
\includegraphics[width=15cm]{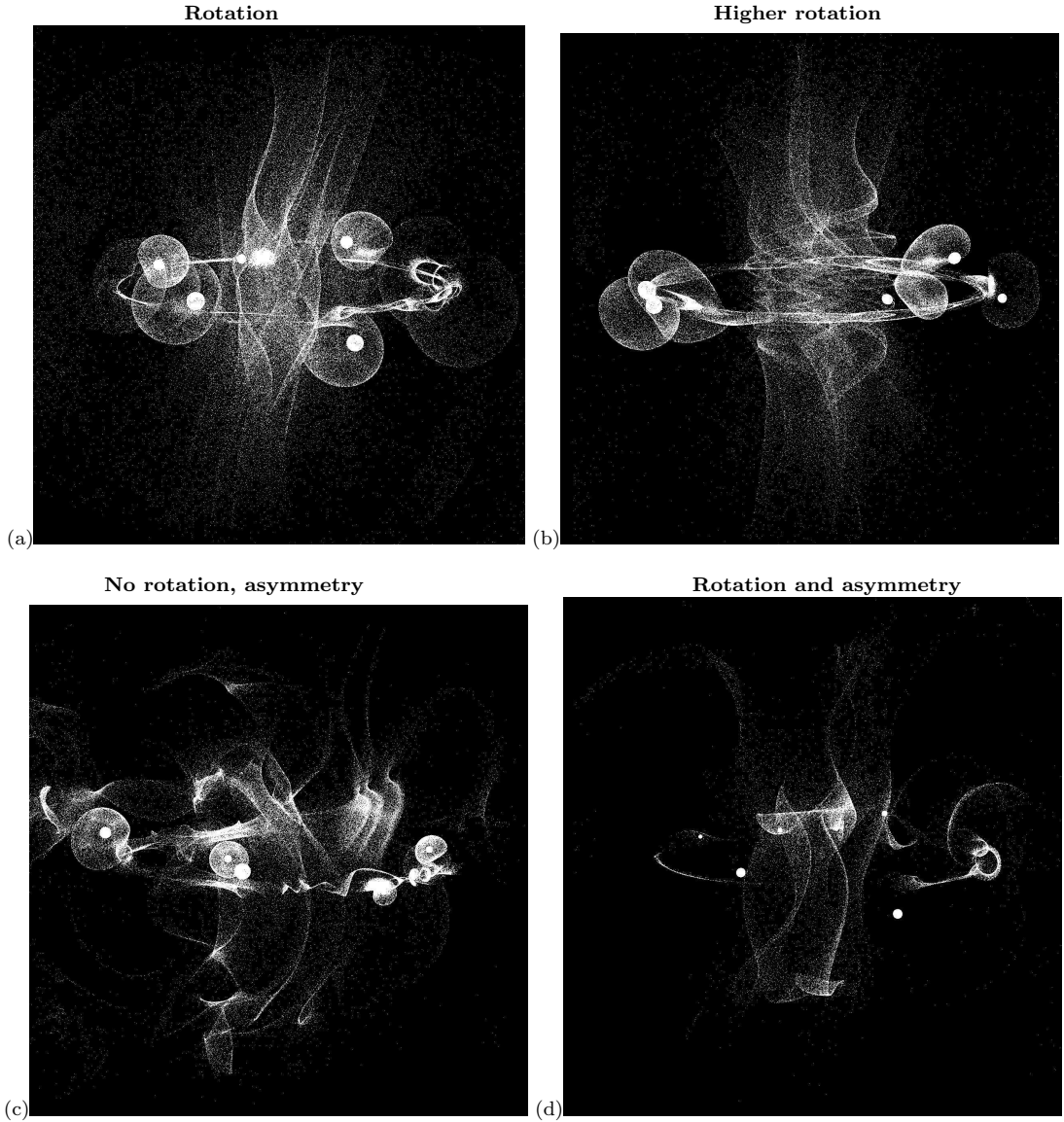} 
\end{center}
\caption{(a) MHD fronts after 5 SNe exploded at random in Galactic rotation.
(b) Same, but the rotation is flat till the center, so that the differential rotation is stronger. 
(c) No rotation, but 3D sinusoidal asymmetry with amplitude $\times 0.5$ and wavelength of $\Lambda=1.2\lu$ is given to the \alf velocity distribution.
(d) Both the mild rotation and weaker amplitude sinusoidal asymmetry ($\times 0.2$, $\Lambda=1\lu$) are given.  Note: (a)-(d) The area sizes are all $3\lu\times 3\lu$.
} 
\label{rotation}  
	\end{figure*}

So far we have neglected the Galactic rotation.
If the disc and halo are rotating, it causes differential rotation on the order of $\sim 1500$ \kms kpc$^{-1}$ in the CMZ, which will twist the wave fronts.
Panel (a) of Fig. \ref{rotation} shows a simulated result, where the surrounding medium is rotating at $\vrot={\vrot}_0 (\varpi/\varpi_0)^2/(1+(\varpi/\varpi_0)^2)$ with ${\vrot}^0=0.07 \vu \sim 150$ \kms and $\varpi_0\sim 0.25 \lu$ ($\sim 25$) pc, approximately mimicking the observed Galactic rotation in the GC{ \cite{so2013}}.  
The observed curved GCFs are well reproduced in this simulation. 
The differential rotation causes twist of the fronts, and the filaments appear bent and spiraling with overlapped filaments crossing each other.
Such crossing and twisted filaments are indeed observed for many MeerKAT filaments{ \cite{yu+2022a}}. 
Rapider rotation causes stronger twist and more complicated filaments as shown in panel (c) where a flat rotation as high as $\sim 200$ \kms is assumed.
These simulations result in more twisted and chaotic filaments than observed, so the the real rotation in the GC would be milder than that assumed here.

\subsection{Asymmetry and turbulence} 

Asymmetry and turbulence in the ISM distribution causes local variations of the \alf velocity, and the waves are refracted and reflected in a more complicated way, yielding more tangled and chaotic front shapes. 
The lower panels of Fig. \ref{rotation} we show such cases with asymmetric \alf velocity distribution.
Panel (c) shows a case without rotation, where a sinusoidal variation is given to the \alf velocity distribution in the form of $V_{\rm A}=V_{\rm A, 0} (1-A\  \Pi_i \sin{2\pi x_i/\Lambda})$ with the amplitude $A=0.5$ and wavelength $\Lambda=1.2\lu$, where $V_{\rm A,0}$ is the unperturbed \alf velocity given in the previous section.
Panel (d) shows the same, but the disc is rotating and the perturbation amplitude is $A=0.2$ and $\Lambda=1\lu$. 
The wave fronts suffer from more chaotic deformation and result in complicated filaments, mimicking some strongly bent and kinked GCFs.
 
 \subsection{{Ring site vs central site of SNe}}

\begin{figure*} 
\begin{center}    
\includegraphics[width=15.5cm]{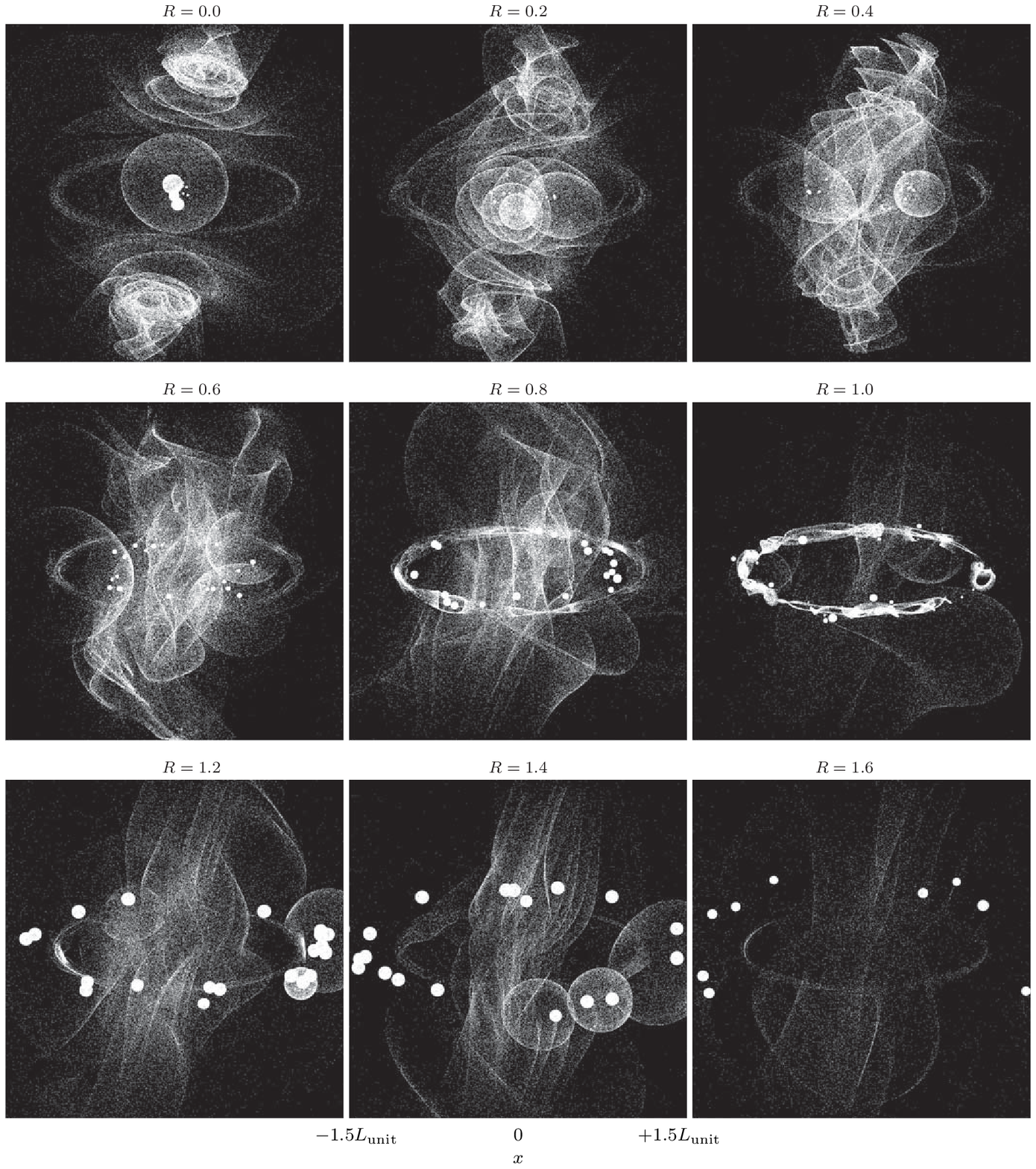} 
\end{center}
\caption{{
MHD wave fronts produced by 20 SNe exploded at random time intervals in the past $20\tu$ ($\sim 1$ My) in different rings of radius $R=0$ (top left) to $R=1.6\lu$ (bottom right) with width of $\delta R=0.1$, as seen from $30\deg$ above the Galactic plane. The area sizes are all $3\lu\times 3\lu$.
The GCFs are produced by SNe exploded at radii $R\gtrsim 0.6-0.8\lu$.
Note that even waves from SNe at such large radii as $R\gtrsim 1.6\lu$ are focused toward the rotation axis, making vertical filaments, though the efficiency decreases.
On the other hand, SNe occurred in the central region at $R\lesssim 0.5$ result in more complicated filaments and loops.
Note also that a significant portion of the waves from SNe exploded inside the SF ring at $R=1.0\lu$ (sixth panel) are trapped by the molecular ring and stay there for longer time, less contributing to the GCF formation.
}}
\label{ring}  
	\end{figure*}

{In order to confirm that the vertical GC filaments are produced by SNe exploded in and near the SF ring, we simulate different cases when the SNe occurred at various radii.
Fig. \ref{ring} shows the results where the SNe exploded in rings of radii $R=0,\ 0.2,\ 0.6,$ ..., $1.6\lu$ with equal width and vertical extent of $\delta R=0.1$ and $\delta z=0.1$.
Explosions of 20 SNe occurred at random time interval in the past $20\tu$ ($\sim 1$ My). 
From the simulation, we find the following properties about the efficiency of filament formation.}

\begin{itemize}
\item {It is shown that SNe exploded in the central region at $R\lesssim 0.5\lu$, which also simulate AGN-origin waves from Sgr A and A$^*$, produce round shaped waves accumulating along the axis of the magnetic cavity filled with the hot plasma, resulting in loops and complicated filaments.
However, these central SNe do not produce the vertical filaments.
The fact that shells or loop filaments are very seldom in the radio maps as in Fig. \ref{mkat10} implies that SN is not a frequent phenomenon in the central $\sim 50$ pc in so far as the presently assumed magnetic structure is correct.
However, it has been also shown that vertical filaments can be produced, if the magnetic cylinder is significantly offset from the rotation axis \cite{so2020th}.
So, the SN rate in the nuclear region remains open as a question.}

\item {SNe exploded in the rings at $R\gtrsim 0.6\lu$ produce vertically elongated filaments.
Even waves from far outside the SF ring at $R\gtrsim 1.6\lu$ are focused around the axis of magnetic cylinder, producing coherent and vertically stretched filaments.
Obviously, the efficiency of focusing decreases at $R\gtrsim 1.6\lu$ because of the wave dilution according to the increasing distance from the GC.
From these simulations, we may conclude that the GCFs are a mixture of old rSNRs originating in a wide ringed area at $R\sim 0.6\lu ~ (\sim 80$ pc) to $\sim 1.6\lu ~ (\sim 200$ pc), predominantly associated with the SF ring in the CMZ.}

\item {It is point out that the waves from SNe exploded near the ring center at $R\simeq 1\lu$ (sixth panel of Fig. \ref{ring}) are trapped by the molecular ring due to the low \alf velocity (Fig. \ref{magrho}), and propagate along the ring without forming vertical filaments.
Therefore, SNe occurred in the ring center, or the "buried SNe" in the dense molecular clouds in the CMZ little or less contribute to the GCF formation.}

\item {
The present simulation on the radius dependence may put more strict correlation between the SN rate and SF rate inferred from the rSNR counting in the sense that the currently estimated SN rate in the previous subsections would have been under-estimated.
They will also put another constraint on the SN rate in the nuclear region in and around Sgr A as well as the nuclear activity.
These will be discussed in a separate paper on SNRs in the CMZ.}
\end{itemize}


\subsection{The SN orchestra plays the grand harp}

\begin{figure*} 
\begin{center}     
\vskip -3mm
\includegraphics[width=10cm]{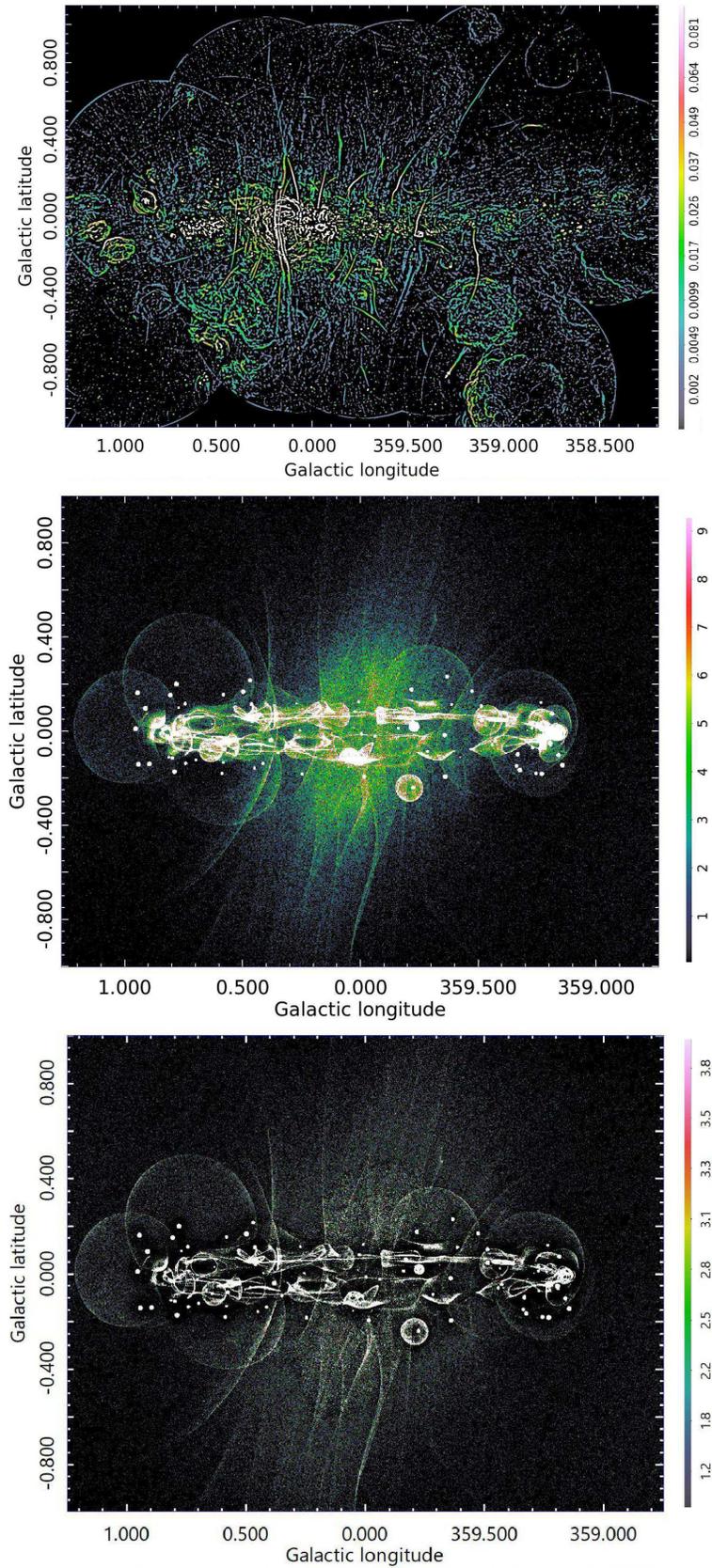} 
\end{center} 
\caption{[Top] MeerKAT 1.28 GHz image from archival data (Heywood et al. 2022) with structures greater than 10 pixels subtracted by BGF(Sofue and Reich 1979).   Intensity scale is mJy per $4''$ beam. 
[Middle] Side view of the MHD wave fronts, where 100 rSNRs exploded at random in the past 0.5 My near the SF ring at inclination $85\deg$. Coordinates are scaled to GC distance of 8.2 kpc. 
[Bottom] Same, but extended features are subtracted for comparison with the BGF image.
} 
\label{mkat10}  
\end{figure*}

Fig. \ref{mkat10} shows the 1.28 GHz radio continuum map made from the MeerKAT data{ \cite{hey+2022}}, where extended components with scale sizes greater than 10 pixels have been subtracted by the background-filtering technique{ \cite{so+1979}}.
The figure also shows the result of MHD simulation of the SNR model, where 100 SNe exploded at random over the past 0.45 My near the SF ring of radius 100 pc and inclination $85\deg$.
This result simulates a case of continuous supernova explosions at an SN rate of $2\times 10^{-4}$ y$^{-1}$ in the CMZ, assuming that relics of SNRs survive for $\sim 0.45$ My and then dissipates.

The middle panel is the $(x,z)$ projection of the wave fronts to be compared with the MeerKAT image in the top panel. The coordinates are scaled to the GC distance of 8.2 kpc.
The bottom panel shows the same, but extended features are subtracted in order for a better comparison with the interferometric observations that are less sensitive to more extended emissions than those corresponding to the shortest baseline ($\sim 10' \sim 23$ pc).
Each filament or a shell represents each front of rSNR at the present time, and hence, the entire filaments represent the "present" shapes of the 100 rSNRs that exploded in the past 0.45 My at random.
White dots represent the initial positions of all the SNe with their radius being equal to the local \alf velocity times 0.05$\tu$.

{We then simulate a more realistic case, where SF activity occurred intermittently in the past $\sim 1$ My, and each SF region was associated with a cluster of SNe exploding at random in a shorter time scale.
The star formation occurs intermittently in localized molecular clouds randomly distributed in the molecular ring with radius $\sim 100$ pc of the CMZ, which composes the SF ring of the same radius \cite{hen+2022,so2022}.
Here, 10 SF regions are born at random interval in the past $10-20\tu$.
Each SF region continues to form stars for a duration of $0.05-0.1\tu$.
Several to 15 supernovae explode at random time interval within this duration in each SF region.}

\begin{figure} 
\begin{center}   
\includegraphics[width=7.5cm]{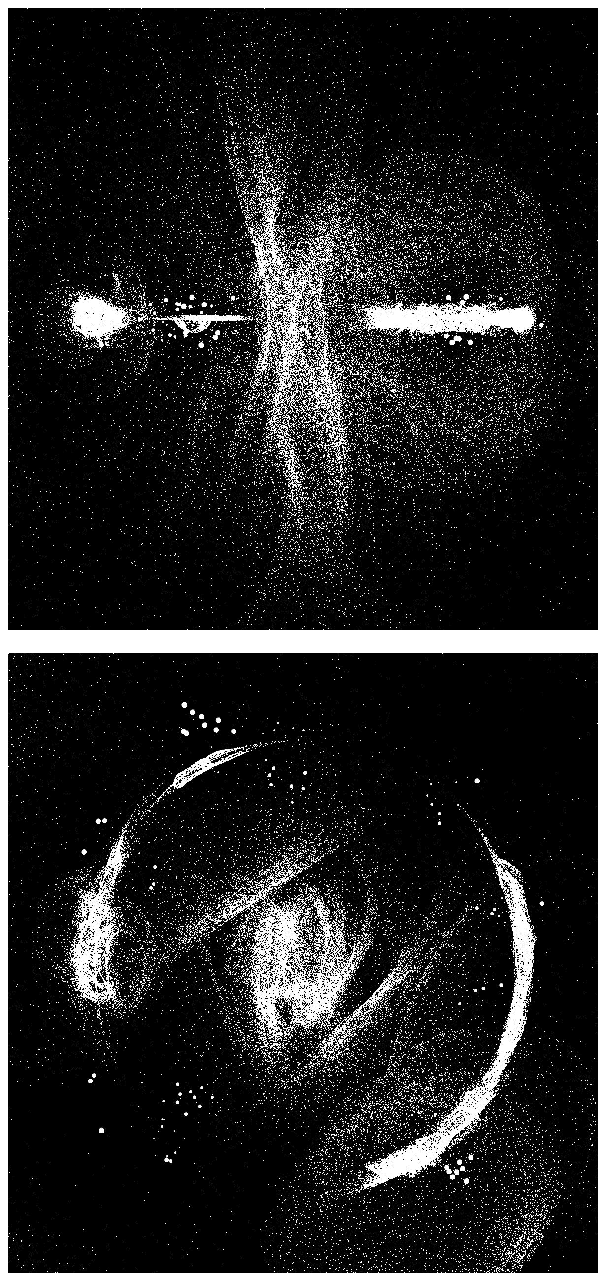}  
\end{center}
\caption{{[Top] MHD fronts projected on the $(y,z)$ plane for the same model as in Fig \ref{mkat2}.
[Bottom] Same but on the $(y,x)$ plane. }}
\label{sfrr}  
	\end{figure}

\begin{figure*} 
\begin{center}  
\vskip -2mm
\includegraphics[width=12.5cm]{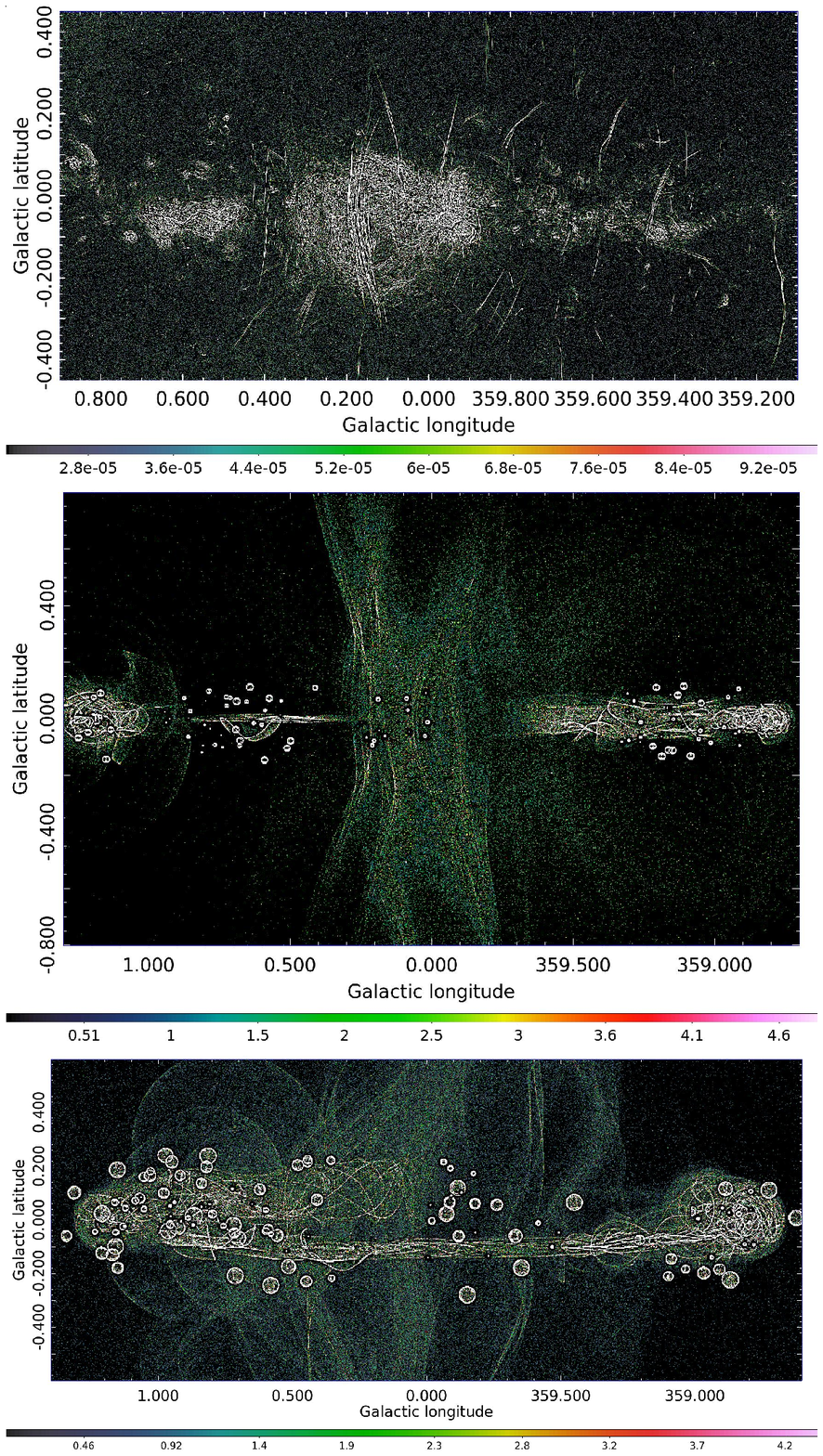} 
\end{center} 
\caption{{
[Top] Same as Fig. \ref{mkat10}, but for the central region after BGF (2 pix).
[Middle] MHD wave simulation of rSNRs in the past $10\tu$, during which 10 SF regions were born, each associated with $5-15$ SNe in $\sim 0.5\tu$.
[Bottom] Same, but a result for a longer duration (past $20\tu$) and time interval for SNe in a wider SF ring.}}
\label{mkat2}  
	\end{figure*}

{Fig. \ref{sfrr} shows the simulated results as projected on the $(x,z)$ (top) and $(x,y)$ (bottom) planes.
Fig. \ref{mkat2} shows the same at higher resolution and compares with the 1.3 GHz MeerKAT image, where extended structures larger than 2 pixles are subtracted in order to enhance finer GCFs.
The bottom panel of Fig. \ref{mkat2} shows a case of a similar intermittent SN and SF activities, but for a different initial condition that the distribution of the SF regions are broader and the duration is about twice up to $20\tu$ in the past.
White circles show the SN positions with the radius being $0.05\tu$.
Due to the longer duration of SF and hence due to older and more expanded fronts (SNRs) than those in the middle panel, the simulated GCFs are more widely distributed.}

{Thus, we have shown that the SNR model can well reproduce the morphological structures of the GCFs, not only the broadly distributed filaments composing the big harp, but also the bunched thin filaments, or the mini harps that make up the entire big harp.}

\subsection{SN ensembles play mini harps: the Radio Arc by a starburst}

\begin{figure*} 
\begin{center}      
\includegraphics[width=15cm]{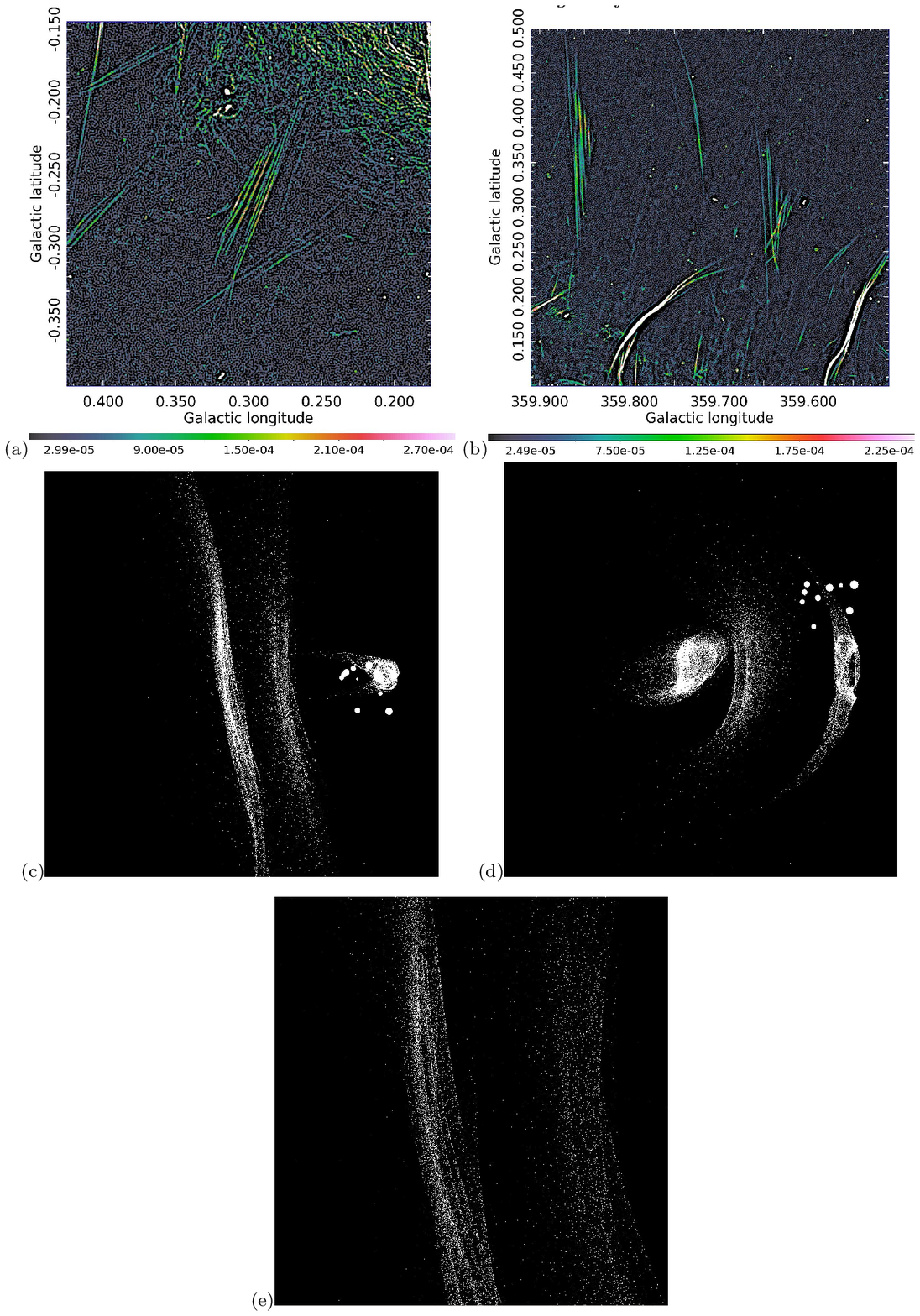} 
\end{center}
\caption{(a) "Harp" G+0.30-0.27, and
(b) "flying mini harps",
as revealed in the 1.28 GHz map with extended structures subtracted as produced from the MeerKAT archival data (Heywood et al. 2022). 
(c)Simulation of the mini harp by MHD waves originating from a cluster of ten SNe exploded $8-9\tu$ ago in an SF region as marked by the small front shells at $t=0.05\tu$. 
Projection on the $(x,z)$ plane ($3\lu\times 3\lu$);
(d) projection on the $(x,y)$ projection.
(e) Same, but close up. 
} 
\label{harp} 
\end{figure*}

So far, we discussed the general morphology of the GCF over the GC region as in Fig. \ref{mkat10}. 
However, closer inspection of the radio images reveals that most of the brightest filaments are consisted of bundles of multiple threads{ \cite{hey+2022,yu+2022b}}, shaped like "mini-harps", as shown in Fig. \ref{harp}.
The well known Radio Arcs can be categorized in the "most prominent mini harp". 
It is stressed that the separation between the filaments is regulated around $\sim 5-20''${ \cite{hey+2022,yu+2022b}}.
The cross section of the typical harp G+0.30-0.26, as inserted in Fig. \ref{harp}, shows 7 filaments spanning for $70''$ with the mean separation of $\Delta L\sim 0.4$ pc. 

\begin{figure*} 
\begin{center} 
\includegraphics[width=14.5cm]{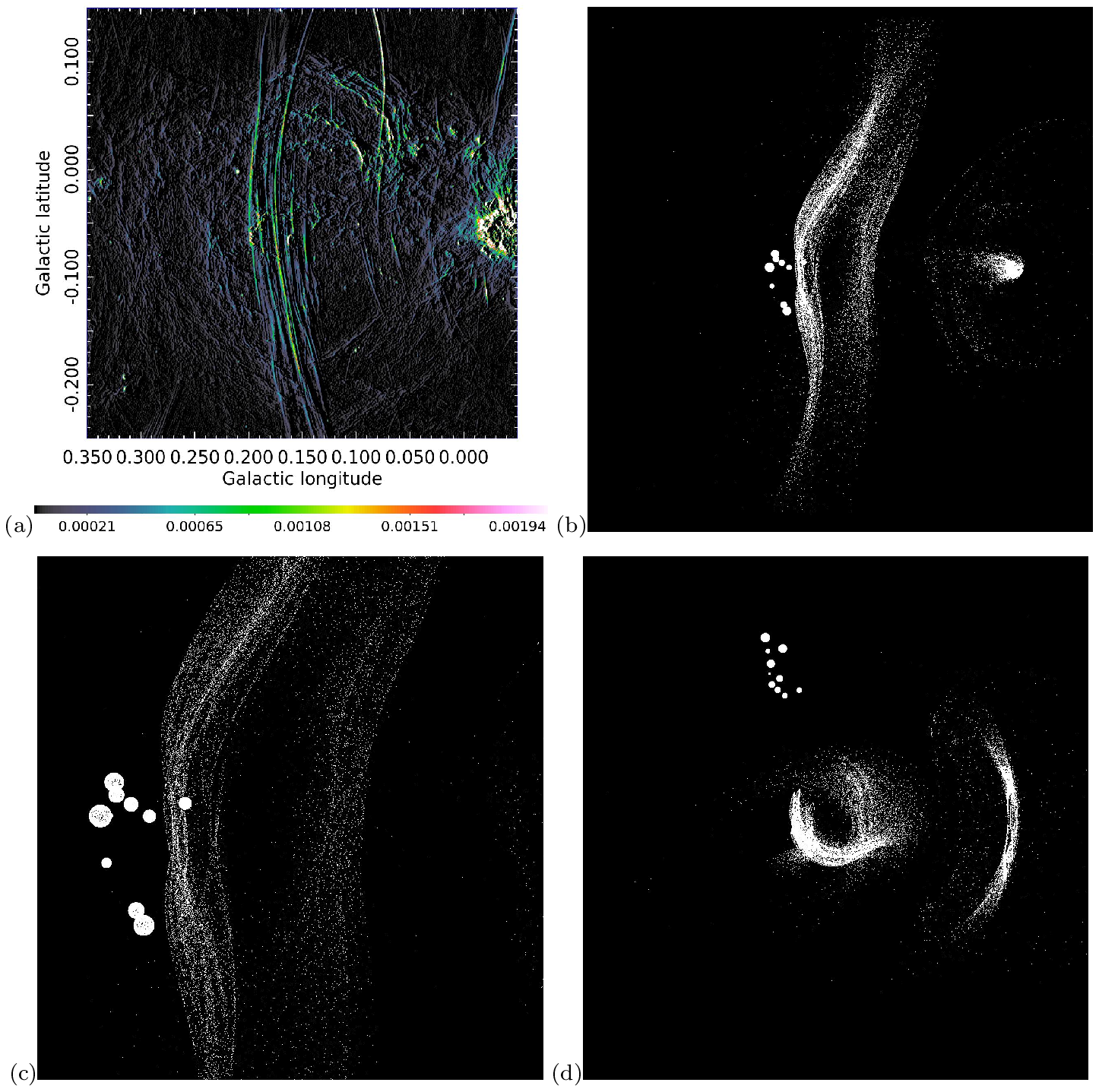}  
\end{center}
\caption{ 
(a) Radio Arc from MeerKAT 1.28 GHz image (Heywood et al. 2022), where extended structures are subtracted. The prominent harp feature suggests that the structure  is due to an SN cluster associated with a recent (within $\sim 1$ My) starburst in the CMZ.
(b) MHD wave simulation of the Radio Arc $(x,z)$ plane, 
(c) same, but close up, and 
(d) projection on the $(x,y)$ plane. } 
\label{harpmhd} 
\label{radioarc} 
	\end{figure*}

Since the propagation speed of MHD waves is $v\sim 200$ \kms (Fig. \ref{magrho}), the separation corresponds to the time interval of $\Delta t=\Delta L/v \sim 2000$ y.
So, the harp is mimicked by the remnants of clustered 7 SNe exploded within $1.4\times 10^4$ y in the SF ring $\sim 0.5$ My ago.
The lower panels of Fig. \ref{harp} show a result of MHD simulation, where a cluster of 10 rSNRs exploded $8-9\tu$ ago at random in a region of radius $0.2\lu$ on the SF ring.
The fronts reasonably reproduce the observed property of the harps.
In the same context, the Radio Arc may be a relic of a cluster of about 15 SNe that took place in the CMZ some $\sim 0.5$ My ago.
The variety of lengths and strength of the harps may be due to the different epochs of explosion and the paths in the ISM with various conditions of wave divergence.
The particularly bright filaments and their high density in the Radio Arc indicate that the responsible SN cluster was more powerful than the other clusters driving the other mini harps and filaments.
This suggests that the Radio Arc may be a relic of recent (within $\sim 1$ My) starburst activity in the CMZ.

The lower panels of Fig. \ref{radioarc} shows a result when a mini starburst took place in an SF region on the SF ring, where ten SNe exploded within a duration of $1\tu$ (0.045 My) about $10\tu$ (0.45 My) ago.
White dots represent the cluster of the SNe with the front shells at $t=0.05\tu$.
The waves are stretched in the $z$ direction, are focused and converged toward the axis, and form a bunch of vertical and twisted filaments on the sky, mimicking the Radio Arc's filaments.

\subsection{{Loop filament: still a shell}}

\begin{figure} 
\begin{center}     \includegraphics[width=7.5cm]{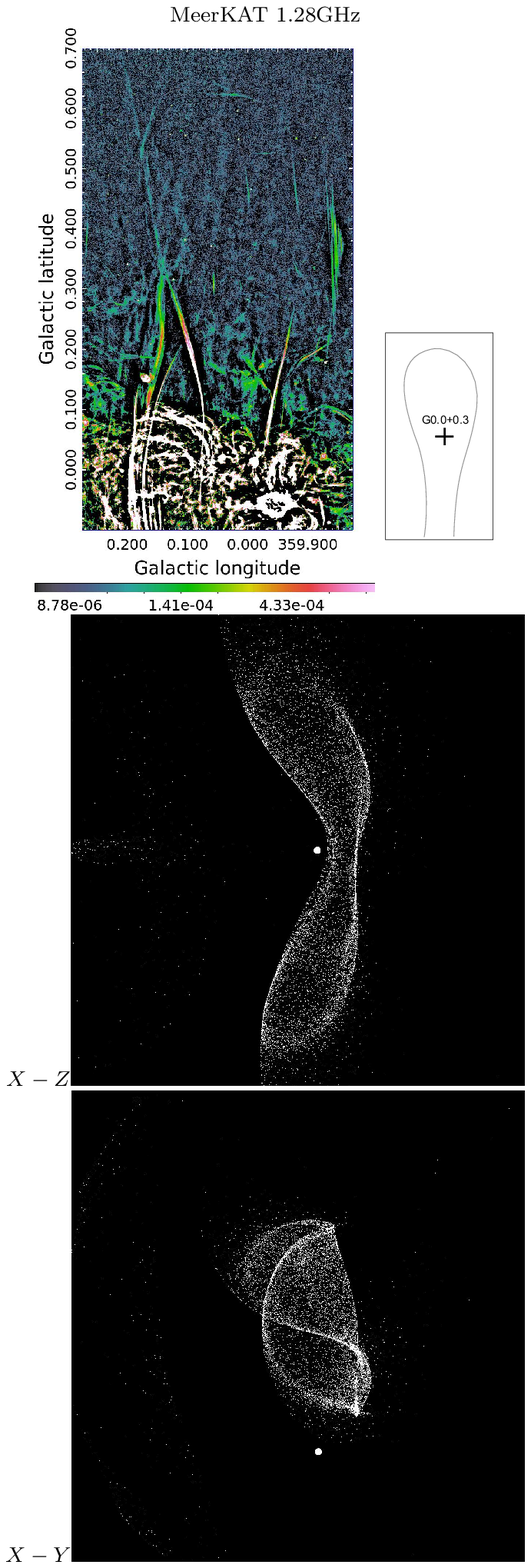}  
\end{center}
\caption{{[Top] Loop filament G+0.0+0.3 at 1.28 GHz map (from MeerKAT archive), drawing a loop as illustrated by the finding chart. 
[Middle] MHD wave front projected on the $(x,z)$ plane at $t=8\tu$ due to an SN exploded at $r=0.5\lu$ inside CMZ. Frame size is $2\lu\times 2\lu$. 
The front is largely stretched in the $z$ direction, and looped at the both tops.
[Bottom] Same, but on the $(x,y)$ plane.}}
\label{loop}  
	\end{figure}

{One of the types of GCFs that are not noticed so far is the "loop filament".
A typical loop is found centered on G+0.0+0.30 as shown in Fig. \ref{loop}, which draws an Omega of longitude and latitude extents of $\Delta l \times \Delta b=0\deg.25 \times 0\deg.6=36 \epc \times 85 \epc$ with two vertical filaments running through $\sim$G+0.10+0.2 and G-0.05+0.15.
The northern top is capped by a horizontal curved filament at G0.0+0.61.
The root of the $\Omega$ extend toward the negative latitudes.
A similar morphology has been recognized as the "chimneys" which are composed of the northern and southern GCLs \cite{hey+2019}, drawing a pair of dumbbell-shaped loops.}

{The loop filament can be understood by a projection of a closed front of the MHD wave due to an SN, or a compact cluster of SNe, exploded inside the SF ring assumed here.
In the bottom panel of the figure, we show a calculated filament as a result of MHD front $\sim 8\tu$ after an SN explosion at $\rad=0.5$ in the Galactic plane.
It is stressed that a closed front is obtained only when the SN explosion takes place at $\rad\sim 0.4 \sim 0.u\lu$, while SNe exploded inside or outside these radii cannot produce such an $\Omega$ shape.
} 
	
\subsection{{Right-angle and crossing filaments}}

\begin{figure} 
\begin{center}  
\includegraphics[width=7cm]{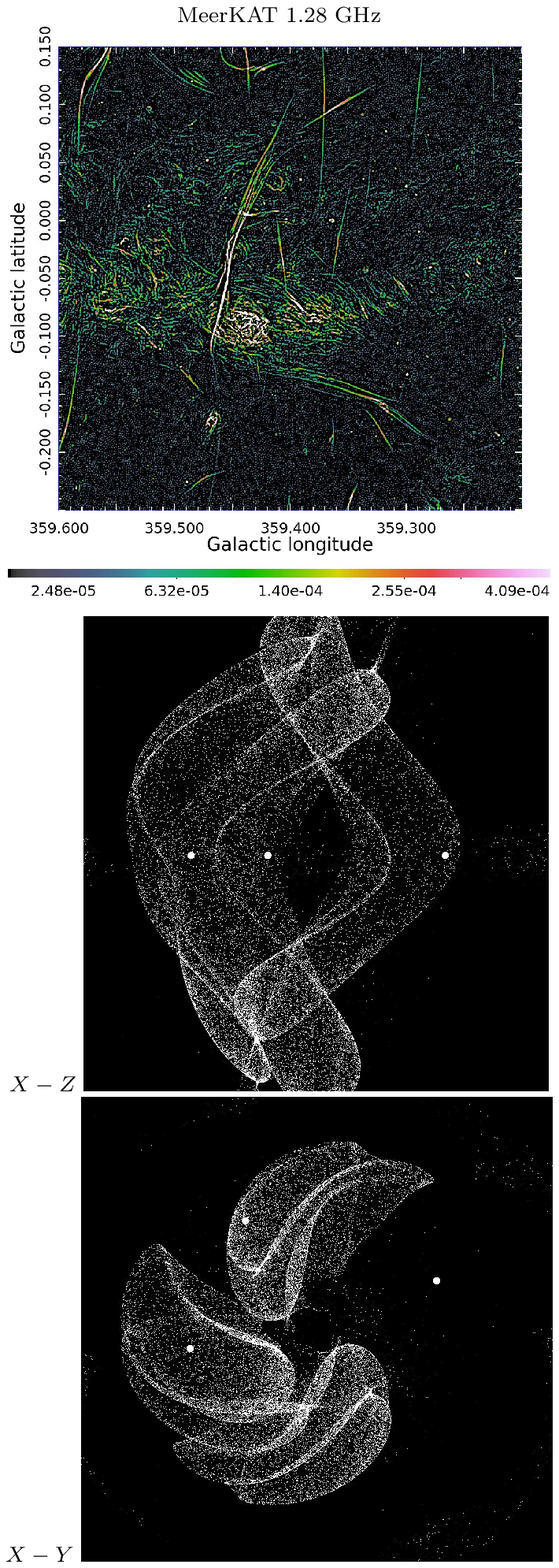}  
\end{center}
\caption{{[Top][ Perpendicular filaments crossing at G-0.55-0.1 in the BGF2 image. They are apparently running near Sgr C at G-0.56-0.08.
[Middle] $(x,z)$ projection of simulated MHD wave fronts at $t=10\tu$ due to 3 SNe exploded at $r=0.5\lu$ inside CMZ. 
The fronts are largely deformed, some are looped, and some are crossing perpendicularly.
[Bottom] Same, but on the $(x,y)$ plane.}}
\label{perp}  
	\end{figure}

{Among the many filaments crossing each other, some filaments intersect the other at large angles.
The most typical case is seen at G-0.56-0.08, where a bright vertical filament near Sgr C is intersected by a horizontal filament, as shown in Fig. \ref{perp}.
The crossing features can be understood as projections of two filaments at different depths overlapped on the line of sight.
The bottom panel of Fig. \ref{perp} shows an example of MHD simulation of such overlapping twisted filaments on the line of sight, where three SNe exploded $10\tu$ ago at $\rad=0.5\lu$.
Some filaments are crossing each other at right angles.}

\subsection{{"Spaghetti": trapped rSRNs in CMZ}}

The simulation shows that MHD waves captured by the molecular ring propagate along the ring, and exhibit tangled noodle-like features near the tangential directions as shown in Fig. \ref{mkat10} and \ref{mkat2}.
Such features have been noticed as the "spaghetti-like" filaments in the MeerKAT 1.3 GHz high-pass images \cite{yu+2022a}. 
In Fig. \ref{spaghetti} we enlarge the tangential region around $l\sim 359\deg$ from the MeerKAT BGF10 image and the MHD simulation. 
The simulation well reproduces the tangled filaments as the wave fronts that are trapped in the low-$\va$ molecular ring, exhibiting complicated helical structures.

However, the observed filaments near the galactic plane may be a mixture with thermal features like HII regions associated with the SF activity. 
So, the model here explains only a part of the phenomenon, and a more sophisticated modeling would be necessary for the origin of the spaghetti filaments. 
A more general study of tangled filaments in the CMZ as well as in the Galactic disc and spiral arms filled with rSNRs would be a subject for the future. 

\subsection{{Fluttering filaments: a wind from GC}}

\begin{figure} 
\begin{center}    
\hskip -5mm\includegraphics[width=8.5cm]{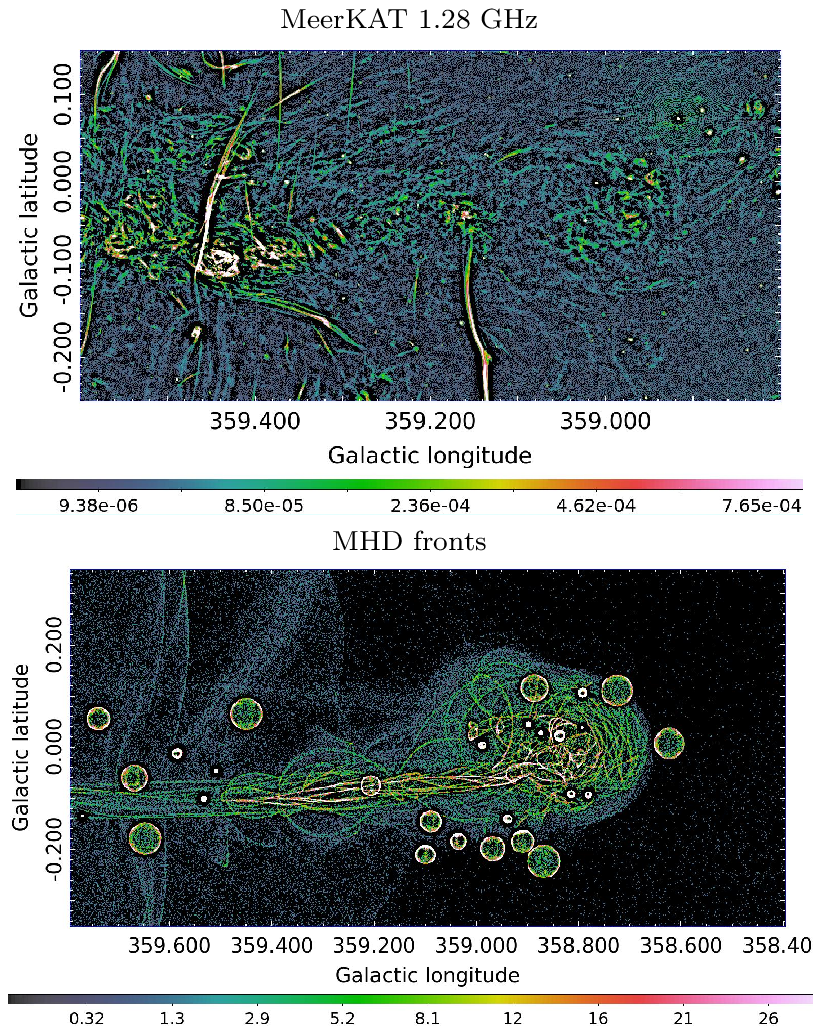} 
\end{center}
\caption{{[Top] 1.28 GHZ BGF map near G359+0.0 from MeerKAT archive (Heywood et al. 2022), showing the "spaghetti" in the tangential region of the warping SF ring (Sofue 2022). 
Numerous filaments are tangling in the ring and are fluttering toward the west. 
[Bottom] Tangential projection of MHD wave fronts trapped by the molecular ring, mimicking the spaghetti. }}
\label{spaghetti}  
	\end{figure}

We also point out that the observed radio GCFs near the galactic plane at $l\sim -0\deg.5$ to $-1\deg.5$ in Fig. \ref{spaghetti} are systematically fluttering toward the west and run almost horizontally.
Interesting, the fluttering phenomenon is lopsided, appearing only in the negative longitude side of GC, but not observed at positive longitudes.
Such fluttering is not explained by the present SNR model, and is not seen in the simulation as shown in Fig. \ref{spaghetti}.
A possible mechanism for causing such fluttering would be a radial flow or expansion of gas from inside the SF ring or a nuclear wind from Sgr A that pushes and blows down the vertical filaments.
Alternatively, they may be blown segments of filaments by the wind which are being caught by the molecular ring. 
In either case, the east-west lopsidedness remains also as a question.
	
\section{Discussion}

\subsection{{The limitation of model}}

{We simulated the evolution of rSNR (relic of SNR) as fully evolved expanded supernova remnants, and mimicked them by heavily deformed MHD disturbances.
The disturbance is approximated by the fast-mode compression MHD wave in the magnetized hot plasma around the GC where the low-plasma $\beta=(c_{\rm s}/\va)^2<1$ condition is satisfied, and the wave propagates at the \alf velocity.}

{The linear wave theory is a powerful tool for finding such new types of phenomena as the cylindrical focusing of the MHD disturbances from the SF ring and Galactic disk onto the rotation axis of the Galaxy.
On the other hand, this method cannot handle the actual magnetic field compression and cosmic ray acceleration required to calculate the radio brightness.
Therefore, we have so far limited our discussion to the morphological properties of the GCF.
Nonetheless, the simulations suggest many aspects about the GCF to consider, and below we attempt a qualitative and speculative discussion of the potentially nonlinear phenomena expected from the model.}

\subsection{Magnetic fields} 

In our previous paper, we showed that the wave model eases the magnetic strength necessary for the radio synchrotron emission by a factor of three  compared to the string model \cite{so2020th}.
From detailed analyses based on the assumption of the pressure equilibrium between magnetic and cosmic-rays, the mean field strength of GCF has been shown to range from 0.1 to 0.4 mG{ \cite{yu+2022a}}. 
If the present wave sheet model is assumed, the field strength would be decreased to 0.03 to 0.13 mG and the magnetic pressure by an order of magnitude, which also eases the acceleration problem of cosmic rays.
    
The Eikonal equations used to trace the MHD waves do not include the magnetic configuration or direction, as the wave propagation depends only on the distribution of the scalar value of \alf speed. 
Even if the ambient fields are oblique and not parallel to the filaments, the magnetic fields in the compressed front will become locally parallel to the wave front, when they emit the synchrotron radiation. 
Namely, the linear polarization observed toward the filaments \cite{la+1999b,ts+1986,so+1987} would represent locally compressed field direction, but does not necessarily represent the background  (undisturbed) field direction. 
In fact, the large and strongly varying Faraday rotation measures observed toward the Radio Arc and some bright filament \cite{ts+1986,so+1987,la+1999a} indicate that the magnetic fields are significantly inclined and twisted from the $z$ axis direction. 
 
Considering the origin of GC magnetic field (Fig. \ref{magview}), the accumulated vertical magnetic field in the GC must be strongly twisted by the differential rotation of the disc and halo \cite{so+2010}. 
However, the GCFs are not observed to be twisted so strongly, as shown in Fig. \ref{mkat10} and \ref{mkat2}.
This fact may manifest such situation that the directions of the undisturbed field and the wavefront are not necessarily be aligned, as assumed in the present MHD-wave model.
Namely, the observed GCFs may not strictly represent the Galactic magnetic field structure.

{Here, we recall that the equations used in the present model do not include the magnetic vectors.
This means that the model does not necessarily require a vertical field orientation as illustrated in Fig. \ref{magview}.
That is, if the \va distribution is fixed, any field, including twisted ones as expected for magnetic towers \cite{kato+2004}, or even spiral and ring fields are equally allowed and the same result is obtained.
In other words, the present model cannot give constraint on the magnetic orientation, although a vertical configuration looks natural in the simple view of the MeerKAT image inside the GC.}

\subsection{Unique morphology in favor of the sheet model}

The present SNR model explains not only the vertical filaments, but also many other filaments and threads having peculiar morphology{ \cite{an+1991,la+1999a,pa+2019,yu+2021,yu+2022a}}, which are attributed to various types of projection of the wave fronts. 
They include the bent radio Arc, horizontal, bifurcated and crossing threads as due to projection of multiple fronts on the line of sight. 
The tightly bunched filaments of the Radio Arc and  the "flying harps" (Fig. \ref{harp}) are simulated by MHD wave fronts due to a cluster of multiple SNe exploded in an SF region. 
Also, threads having kinks, knots, and mouse-like spots can be explained by superposition of obliquely corrugated or twisted fronts{ \cite{so2020th}}. 
We also stress that the similarity between the GCFs and the thin filaments in the Cygnus Loop{ \cite{blair+2005}} would support the SNR model.
Furthermore, if they were real strings, some of the thousands of GCFs would be observed almost parallel to the line-of-sight and appear as extremely bright point-like sources.
The fact that such bright source is not observed would be in favor of the sheet model.

\subsection{Energetics and SN rate} 

The magnetic energy contained in a single thread is estimated to be 
\be{}E\sim (B^2/8 \pi) a_1 a_2 a_3 \sim 4\times 10^{47} {\rm ergs},\ee
where $B\sim 0.1$ mG, $a_1\sim 0.1$ pc is the filament thickness, $a_2\sim 70$ pc ($\sim 0\deg.5$ for the long threads) is the length, and $a_3 \sim 4$ pc is the depth on the line of sight for the thickness $a_1$ and curvature $a_2$.
Therefore, only $\sim 0.1$\% of the released kinetic energy by a single core-collapse supernova (SN)  ($10^{51}$ erg) is sufficient to drive one thread.

Using the background-filtered MeerKAT map shown in Fig. \ref{mkat10} and \ref{mkat2}, we roughly count $\sim 10^2$ filaments inside the supposed magnetic cylinder at $|l|\le \sim 0\deg.5$ and $|b|\le \sim 1\deg$. Detailed statistics of the filaments reports an order of greater number, $\sim 10^3$, including fainter filaments {\cite{yu+2022a}}. 

So, the total energy possessed by the entire filaments is on the order of $\sim 10^{50-51}$ erg, or one SN is sufficient to supply the entire energy.
However, considering the dissipation and dilution of the waves, a larger number of SNe, or about $\sim 100$ SNe in the present model, may have been necessary (and sufficient) to excite the observed filaments.

The current SF rate in the CMZ has been estimated to be $R_{\rm SF}\sim 0.07\Msun$ y$^{-1}$ \cite{hen+2022}.
The SN efficiency per born-stellar mass is estimated by 
$
\eta_{\rm SN}=\int_{10\Msun} ^\infty \mathcal{M}(m) dm/\int_{0.1\Msun} ^{10\Msun} \mathcal{M}(m) dm\sim 1.9\times 10^{-3},
$
which is equal to the high-mass stars' mass density over the mass density of all stars with $\mathcal{M}(m)\propto m^{-2.36}$ being the mass function of the stellar mass $m$.
Combining with the SF rate, this SN fraction yields an SN rate of $R_{\rm SN}\sim 1.3\times 10^{-4}$ y$^{-1}$ in the CMZ. 
This rate agrees with the SN rate ($2\times 10^{-4}$ y$^{-1}$) assumed in the simulation in Fig. \ref{mkat10} within a factor of two.
The agreement between the counting of filaments (rSNRs) and the SF rate would be in favor for the SNR model. 

\subsection{{SN History and the high density of SNR}}
     
\begin{figure} 
\begin{center}  
\includegraphics[width=8.5cm]{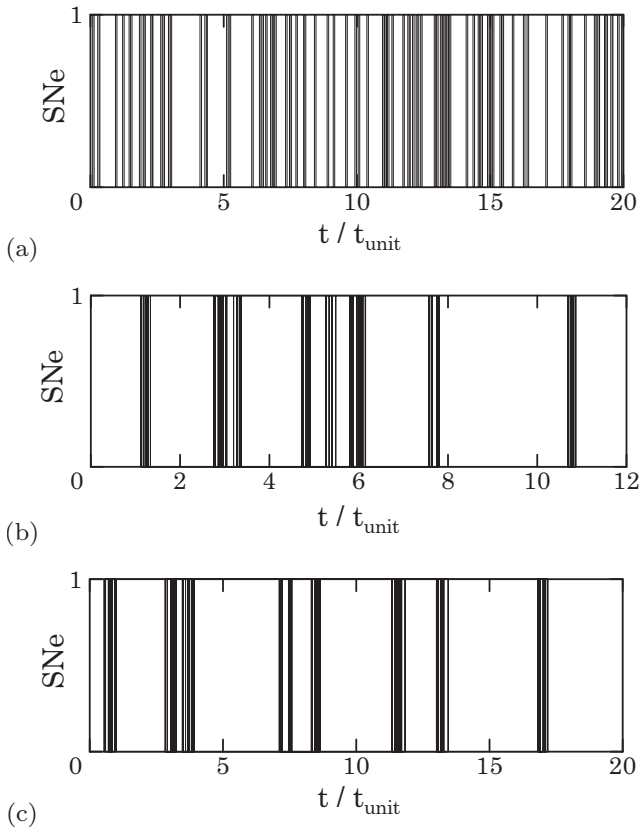}  
\end{center}
\caption{{  
(a) Random occurrence of SNe in the past $20\tu \sim 1$ My for MHD simulation of the intermittent SN/SF model used in Fig. \ref{mkat10}).  
(b) Same, but for Fig. \ref{mkat2}, middle panel.
(c) Same, but for Fig. \ref{mkat2}, bottom panel. 
}}
\label{history1}  
	\end{figure}

\begin{figure} 
\begin{center}   
\hskip -5mm\includegraphics[width=9cm]{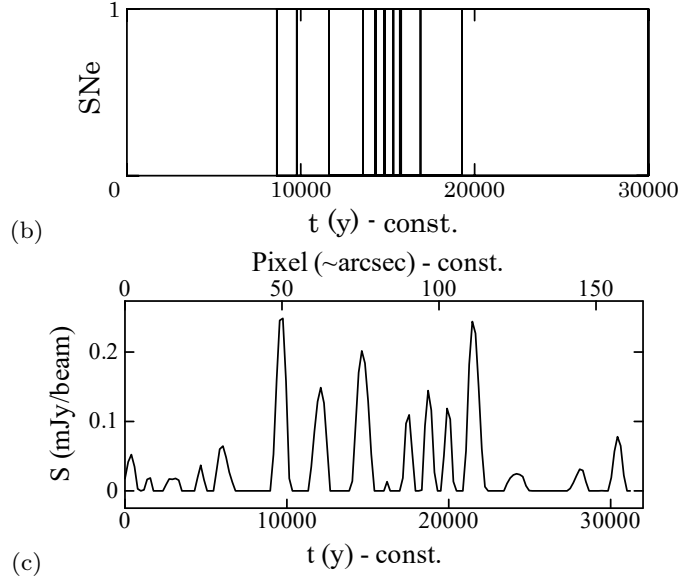} 
\end{center}
\caption{{   
(a) Enlargement of one of the SN clusters from panel (b) of Fig. \ref{history1} with the horizontal axis converted to time. 
(b) Horizontal cross section of Harp G+03-0.2 from Fig. \ref{harp} with the filament separation converted to time in year for an assumed front velocity of 200 \kms. 
}}
\label{history2}  
	\end{figure}

{The thus estimated SN rate is consistent with that assumed in the simulations of the SNR model for GCF.
Fig. \ref{history1} shows the frequency of SN explosions in the past $10-20\tu$ as assumed in the intermittent SF/SN model shown in the previous subsection (Fig. \ref{mkat10}, \ref{mkat2}).}
{Fig. \ref{history2} panel (a) enlarges one the SF periods from panel (b) of Fig. \ref{history1}, showing a cluster of $\sim 10$ SNe within $\sim 10^4$ y.
This diagram may be compared with the cross section of the mini harps of GCF.
In panel (b) we show a horizontal cross section of the radio intensity distribution of the mini harp G+0.30-0.26 as read from Fig. \ref{harp}.
By our interpretation each peak corresponds to an rSNR, so that the spatial separation of filaments represents the time interval of SN explosions.
In this diagram, 7 filaments are counted, spanning for $70''$ with the mean separation of $\Delta L\sim 0.4$ pc, so that the profile indicates SN explosions at a mean interval of $\sim 2\times 10^3$ y in a period of $1.4\times 10^4$ y.}

According to the present model the spatial density of rSNRs in the central 100 pc of the GC is as high as $n_{\rm SNR}\sim 10^2 (\pi(100{\rm pc})^2 \times 100{\rm pc})^{-1}\sim 4\times 10^4 {\rm kpc}^{-3}$.
This density may be compared with the density of SNRs in the Galactic disc of $\sim 300-1000 (\pi (10{\rm kpc})^2\times 0.1{\rm kpc})^{-3}\sim 10-30 {\rm kpc}^{-3}$ \cite{green2019,anderson+2017}.

The high (r)SNR density in the GC than that in the disc would be due to several reasons as follows.
\begin{itemize}
\item The intrinsic SN rate per unit volume in the GC is higher than that in the disc.
\item The GC SNRs are accumulated toward the center by the focusing effect of the wave propagation in the magnetized cylinder filled with the plasma.
This effect acts to keep the old SNRs in the restricted area in the GC from diluting away.
\item The disc SNRs have been searched for by assuming a spherical shape. 
In other words, old SNRs that lost their spherical shape are difficult to discover both due to the irregular shape and faintness.
\item Mismatch of the number of SNRs and the SF rate in the Galactic disc \cite{anderson+2017} would be due to such reasons.
\item Therefore, the GC is a special region for the SNR research, because most of the SNRs (rSNR) younger than $\sim 0.5$ My are visible in this restricted area.
This makes a contrast to the current SNR research in the Galactic disc, where the majority is not found yet or not visible.
\item GC is also special, because the (r)SNRs are closely associated with the SF sites and molecular clouds in the small volume of $r\sim 100$ pc area.
This also makes contrast to the rare or almost no spatial correlation of the catalogued SNRs to their parent SF regions in the Galactic disc.
\end{itemize} 
 
\subsection{{SNR-SNR collision }}
    
The GCFs as rSNRs through their dissipation will act as the heating source of the hot plasma with thermal energy on the of the order of $\sim 10^{51-52}$ {\cite{nax+2019,po+2019}}.
The present model offers an efficient energy supply by SNR-SNR collisions (SSC) from the SNe exploded in the SF ring by the focusing of the waves toward the center. 

As discussed in section \ref{secdissipation}, the dissipation length is on the order of several tens to $\sim 100$ pc in the magnetized X-ray hot plasma.
So, the waves are diluted inside the magnetic cylinder, and heat the hot plasma filling the GC lobe {\cite{so+1984}}.
On the other hand, the damping distance is much longer in the molecular ring with higher density.
Therefore, the waves trapped in the SF ring survive for much longer life and propagate along the ring, drawing the complicated patterns as displayed in the simulations shown in Fig. \ref{mkat10}, \ref{mkat2} and \ref{spaghetti}.

The high density of rSNRs (SNR) inside the magnetic cavity causes frequent collision of the wave fronts. The collision frequency per one SNR is estimated to be on the order of $f_{\rm col}=1/t_{\rm col}\sim 4\times 10^{-5}$ y$^{-1}$ for $N\sim 10^2$ SNRs (sheets) moving at $V\sim 500$ \kms with a mean cross section $\sigma_{\rm SNR}\sim 50{\rm pc}\times 50{\rm pc}\sim 2.5\times 10^3{\rm pc}^2$, which yields a mean free path $l_{\rm col}\sim 1/(n_{\rm SNR}\sigma)\sim 13{\rm pc}$ and $t_{\rm col}=l_{\rm col}/V\sim 2.6\times 10^4$ y. So, the total collision frequency in the GC is on the order of $f_{\rm total}=Nf_{\rm col}\sim 4\times 10^{-3}{\rm y}^{-1}$. The frequency is highest near the rotation axis due to the focusing effects from different and even from counter directions. The SSC is therefore quite frequent in the particular circumstance of GC. 
 
In so far as the waves are linear with small amplitude, they penetrate through each other, and are indeed observed as the crossing filaments.
However, when they are focused around the magnetic-cavity center, the fronts will be amplified and become non-linear shock waves. 
The shock wave will cause more effective heating of the plasma as well as the acceleration of cosmic rays.
The acceleration takes place not only by the direct collision, but also by the Mach stem, wiping a wider area with the amplified strength \cite{hartigan+2016,vieu+2020}.  

\subsection{{Radio spectrum}}  

\begin{figure} 
\begin{center}    \includegraphics[width=8cm]{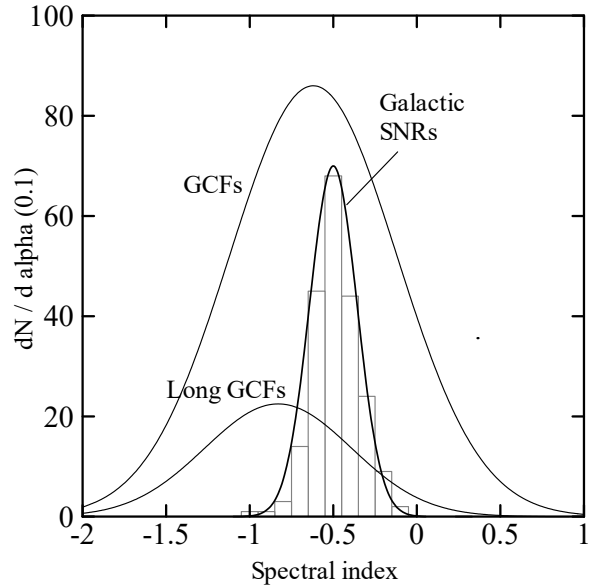} 
\end{center}
\caption{{Number of GCF as a function of spectral index $\alpha$ given by Yusef-Zadeh et al. (2022a) compared with that for Galactic SNRs (histogram and approximate fit by Gaussian profile) as constructed from the SNR catalogue by Green (2019). Plotted values are $dN/d\alpha$ per $\delta \alpha=0.1$.}}
\label{SNRalpha}  
	\end{figure}

{In Fig. \ref{SNRalpha} we reproduce the distribution diagrams of the radio spectral index $\alpha$ ($\Sigma \propto \nu^\alpha$ with $\Sigma$ the radio brightness and $\nu$ the frequency) of GCFs obtained by the large-scale statistical analysis of the MeerKAT data \cite{yu+2022a}.
The GCFs with length $>66''$ have indices of $\alpha=-0.62\pm 0.5$ and long GCF with $>132''$ have steeper index of $-0.82\pm 0.4$. 
We compare them with the distribution of spectra taken from the catalogue of Galactic SNRs \cite{green2019,green2019cat}, which shows $\alpha=-0.5\pm 0.2$. 
Therefore, the spectral indices of GCFs and SNRs coincide with each other within their dispersion, and are consistent with the idea that GCFs are rSNRs.
It is also noticeable that the mean value of $\alpha$ decreases from SNR to GCF and from GCF to long GCF. This is reasonable in view that the longer the filament or the larger the SNR, the older the filament and the steeper the cosmic-ray electron spectrum due to aging.
In this sense, the GCFs are similar to such extended SNRs like Vela X, Y, Z having $\alpha= -0.4 \sim -0.8$ \cite{alva+2001} and old SNR S147 with $\alpha=-0.3 \sim-1.3$ \cite{so+1980,fuerst+1986,xiao+2008}. }

{Another important aspect in Fig. \ref{SNRalpha} is the wider dispersion of $\alpha$ for the GCF compared to that of the SNR. This is understandable for two reasons. One reason is that, as mentioned above, GCF is much older. Another is that the count of old SNRs is missing due to the difficulty of detection at a larger diameter \cite{green1984}. Nevertheless, the question remains why the GCF can have both such steep and mild tails that reach $\alpha \sim -2$ and $+0.5$. This requires extremely sharp energy spectrum of cosmic rays reaching $\beta_{\rm CR}\sim -4$ as well as very flat spectrum reaching to $\sim -1.5$. This problem would be a challenging subject for the acceleration mechanism of cosmic rays in the GC.}

{To summarize, the steeper spectra of GCFs (rSNR) would indicate that they are older than galactic SNRs ($10^2-10^4$ y). This agrees with the result that the simulated GCFs focused around the rotation axis are more aged than $\sim 10^5$ y.
In a similar context, but in terms of the distance from the explosion center instead of age, the observed steepening of $\alpha$ of GCFs with the increasing latitudes \cite{yu+2022a} may be explained by the cooling (instead of aging) effect of the filaments. Namely, the more distant is the filament from the excitation center (SN) in the CMZ, the steeper the spectrum.}

\subsection{Surface brightness-diameter relation}

We examine if the radio brightness of the GCF is consistent with that expected for SNRs using the $\Sigma-D$ (surface brightness-diameter) relation for SNRs. 
In the present model, the surface brightness $\Sigma_{\rm f}$ of a filament represents  tangential integration of a sheet of side-on brightness $\Sigma$.
It is approximately related to the width $w$ of the filament and diameter $D$ (twice the curvature) by $\Sigma\simeq\sqrt{w/D}~ \Sigma_{\rm f}$. 
Inserting this relation to the $\Sigma-D$ relation of SNRs \cite{case+1998}, we obtain $D\sim 150-10$ pc for observed brightness $\Sigma_{\rm f}\sim 0.1-10$ mJy $4''$-beam$^{-1}$, $\alpha\sim -0.6$, and filament thickness of $w\sim 4-10''\sim  0.2-0.4$ pc \cite{yu+2022a}.
However, this estimation may include large uncertainty, because individual filaments are small sections of the largely deformed rSNR.
So, we can only state that the radio brightness of GCFs is not inconsistent with the $\Sigma-D$ relation.

For comparison, we estimate the same for the old shell-type SNR S147.
This SNR is observed to have a total flux of $S_{\rm 1 GHz} \sim 69$ Jy over a round area of diameters of $2.\deg5\times 2\deg.8$ \cite{xiao+2008}.
This leads to the surface brightness of $\Sigma_{\rm 1 GHz}=4.1\times 10^{-22}$w m$^{-2}$ Hz$^{-1}$ str$^{-1}$, and the diameter is estimated to be $D\sim 101$ pc and the distance of $2.2$ kpc from the $\Sigma-D$ relation.
So, the GCF in view of the SNR model is similar to the old shell-type SNR S147 \cite{so+1980,alva+2001}.
We compare the parameters derived for the GCF in the regime of SNR with those observed for S147 in table \ref{tabsnr}. 
 
\begin{table}   
\caption{Approximate parameters of a GCF as an SNR compared with the SNR S147}
\begin{tabular}{lll} 
\hline 
\hline
Object & GCF$^1$ & S147$^{2,3,4}$   \\
\hline
Shape& Filaments & Shell\\ 
$\Sigma$ (surf. brightness) &$0.1-10$ mJy beam$^{-1}$ &$4.1\times 10^{-22}$ $^{\dagger}$ \\
$\alpha$ (spectral index)& $-0.6\sim -0.8$ & $-0.3 \sim -1$ \\
$\Theta$ (ang. dia. degree)& ---& $2.5\times 2.8$ \\
$D$ (dia., pc)& $10-150^{\#}$& 101 $^{(5)}$\\
Distance (kpc)  & (8.2)  &2.2 \\
Age (My) & 0.1-0.5$^\ddagger$ &0.4  ($\sim\frac{2}{3}\frac{D}{2}/V_{\rm expa}$) \\ 
$V_{\rm expa}$ (\kms)& 100-1000$^\ddagger$ & 80 \\
$B$ (mG) & 0.1& 0.02 \\ 
\hline  
\label{tabsnr}
\end{tabular}\\
1) Yusef-Zadeh et al. (2022a); 2)Xiao et al. (2008); 3) Fuerst et al. (1986); 4) Sofue et al. (1980); 5) $\Sigma-D$ relation (Case et al. 1998)\\
$^*$ 1.3 GHz for GCF, and 1 GHz for S147 (w m$^{-2}$ Hz$^{-1}$ str$^{-1}$) \\ 
$^{\dagger}$ Total flux 69 Jy over $2\deg.5\times 2\deg.8$ dia. ellipse.\\ 
$^{\#}$ Twice the curvature for GCF.\\
$\ddagger$ \alf velocity.\\
\end{table} 

\subsection{AGN origin vs SNR origin}

In our previous paper \cite{so2020th} we showed that the GCF can be produced as the result of puffing AGN activity at the nucleus or by SNe near Sgr A, where we assumed a similar, but more off-center oblique cylinder of magnetic field.
In the present paper, we showed that the observed brightness of individual GCF as averaged over the supposed wave area is consistent with the brightness of old SNRs satisfying the $\Sigma-D$ relation.
This means that the source energy of each filament is regulated on the order of that of one SN, $\sim 10^{51}$ ergs, supporting the SNR origin scenario.
On the other hand, the AGN activity hypothesis does not necessarily require such regulation of the released energy. 

As to the location of the SN explosions, either in Sgr A or in CMZ, the latter seems more natural for the following reasons, although we cannot exclude the possibility that some SNe exploded in/near Sgr A are contaminated. 
First, the number of the filaments distributing over the wide area \cite{yu+2022b} is consistent with the present SNR model at the SN rate as expected for the SF rate in the CMZ \cite{hen+2022}.
Second, the surface brightness averaged over the area covered by the supposed line-of-sight depth is consistent with that for an SNR from the $\Sigma-D$ relation.
Third, the filament's intensity is regulated around $\sim $ mJy beam$^{-1}$ within a factor of 10 \cite{yu+2022a}, consistent with the rSNR scenario, but such regulated intensity would be difficult to explain by AGN events.

\subsection{Relation to GC bubbles}

According to the SNR model, the GCFs are objects with life times less than $\sim 0.5$ My and total energy on the order of $\sim 10^{53}$ ergs ($\sim 100$ SNe) during the ordinary phase of SF in the CMZ \cite{hen+2022}.
The outflow phenomena that produced GCL \cite{so+1984,hey+2019} and X-ray hot plasma \cite{nax+2019,po+2019} may have occurred prior to or in parallel to the current SF activity.

On the other hand, the GC experienced more energetic explosion and outflow $\sim 5-20$ My ago with the released energy of $\sim 10^{55}$ ergs \cite{so1977,so2000,bla+2003,so+2016,ka+2018}, as evidenced by the detection of the giant bubbles of radii a few to $\sim 10$ kpc in the radio \cite{haslam+1982}, X-ray \cite{sno+1997,pre+2020} and $\gamma$-ray emissions \cite{su+2010}. 
However, such "ancient" events at $t\gtrsim 5$ My should have no longer direct effect to the "recent" SF activity at $t \lesssim 0.5$ My in the CMZ, because the regulation time scale there is comparable to the Galactic rotation period of a couple of My \cite{sormani+2020}. 

In fact, the GCFs are coherently aligned to each other and are not strongly disturbed.
The globally regular morphology would manifest a rather undisturbed and quiet ISM condition in the recent ($\lesssim 1$ My) CMZ and its surroundings.
The milder outflows such as the GCL and hot plasma of $\sim 100$ pc scale and $\sim 10^{53}$ ergs can be coexisting inside the ring of the CMZ and SF ring in their basic states.

\subsection{Proper motion of the filaments}

Finally, we comment on the proper motion of the filaments.
In the present model, the wave fronts are moving at \Alf velocity of $\sim 100-1000$ \kms.
This would cause proper motion between the overlapping filaments of $\sim 3-30$ mas y$^{-1}$.
Mutual velocities between the overlapped filaments in the near and far sides of the GC would be even higher.
If they are obliquely crossing, the crossing points will move much faster along the filaments. 

Analysis of the mutual positional displacements of the filaments with respect to the neighboring point sources \cite{yu+2022c} at different epochs would be an ideal method to detect the proper motion by regarding the point sources as the reference, which are supposed to be at cosmological distances.

\section{Summary}
 
We have simulated the evolution of rSNRs in the Galactic Center in the presence of a vertical magnetic cylinder filled with hot plasma by tracing the propagation of fast-mode magneto-hydrodynamic (MHD) waves.
The waves are shown to be focused around the Galactic rotation axis, and form cylindrical sheets whose tangential projection on the sky constitutes the vertical filaments.
The GCFs are, therefore, explained as old relic of SNRs that exploded in the SF ring in the past $\sim 10^5$ years at an SN rate of $\sim$ several$\times 10^{-4}$ y$^{-1}$.
The SNR model can explain not only the morphology, but also the steep radio spectrum, the smoothed brightness over the distribution area that satisfies the surface brightness-diameter relation of SNRs, and the heating energy source of the hot plasma in the GC.

We thus conclude that the GCFs are old relic of SNRs similar to the well-known shell-type SNRs such as the Cygnus Loop, except that they are significantly deformed in the particular circumstance of the GC.
According to the SNR model we need not to assume any unknown, uncertain, or sophisticated mechanism such as ejection, magnetic inflation, and/or instabilities to create the magnetic threads.
The only difficulty with this model may be that we have to abandon the common sense that the shell SNRs are spherical.

{The rSNR hypothesis would thus open a new era of an integrated study of the ISM composed of the molecular clouds, star forming regions, and SNR (SNe) that are all visible in the same field of view at the precise distance in the GC, well observed at the same time in radio (from molecular lines to continuum), infrared (lines and continuum), X rays and gamma ray emissions. We therefore emphasize from a new point of view that the GC is a particular place in the Milky Way for the study of the ordinary ISM and its evolution with SF feedback via the rSNRs.}

\vskip 3mm
\noindent {\bf Acknowledgements} 
The computations were carried out at the Astronomy Data Center of the National Astronomical Observatory of Japan.  The author is indebted to the MeerKAT team (Dr. I. Heywood and the collaborators) for the 1.3 GHz radio data. The MeerKAT telescope is operated by the South African Radio Astronomy Observatory.\\

\noindent {\bf Data availability} 
The radio data have been downloaded from  https://archive-gw-1.kat.ac.za/public/repository/10.48479/fyst-hj47/index.html. \\

\noindent {\bf Conflict of interest}
The author declares that there is no conflict of interest.


\begin{thebibliography}{}  

\bibitem[Alvarez et al. 2001]{alva+2001} Alvarez, H., Aparici, J., May, J., et al.\ 2001, \aap, 372, 636.

\bibitem[Anantharamaiah et al. 1991]{an+1991} Anantharamaiah K.~R., et al.
 1991, MNRAS, 249, 262 
 
\bibitem[Anderson et al. 2017]{anderson+2017} Anderson, L.~D., Wang, Y., Bihr, S., et al.\ 2017, \aap, 605, A58. 

\bibitem[Barkov et al. 2019]{ba+2019} Barkov M.~V., Lyutikov M., 2019, MNRAS, 489, L28 
 
\bibitem[Blair et al. 2005]{blair+2005} Blair, W.~P., Sankrit, R., \& Raymond, J.~C.\ 2005, AJ, 129, 2268.

\bibitem[Bland-Hawthorn and Cohen 2003]{bla+2003} Bland-Hawthorn, J. \& Cohen, M.\ 2003, \apj, 582, 246.

\bibitem[Boldyrev et al. 2006]{bo+2006} Boldyrev, S., \& Yusef-Zadeh, F.\ 2006, ApJL, 637, L101
 
\bibitem[Case \& Bhattacharya 1998]{case+1998} Case, G.~L. \& Bhattacharya, D.\ 1998, \apj, 504, 761. 

\bibitem[Dahlburg et al. 2002]{da+2002} Dahlburg, R.~B., et al. 
2002, ApJ, 568, 220
atistics

\bibitem[Fuerst \& Reich 1986]{fuerst+1986} Fuerst, E. \& Reich, W.\ 1986, \aap, 163, 185 

\bibitem[Green 1984]{green1984} Green, D.~A.\ 1984, \mnras, 209, 449.
\bibitem[Green 2019]{green2019cat} Green, D.~A.\ 2019, VizieR Online Data Catalog, VII/284
\bibitem[Green 2019]{green2019} Green, D.~A.\ 2019, Journal of Astrophysics and Astronomy, 40, 36. 

\bibitem[Hartigan et al. 2016]{hartigan+2016} Hartigan, P., Foster, J., Frank, A., et al.\ 2016, \apj, 823, 148.


\bibitem[Hasegawa et al. 1994]{ha+1994} Hasegawa, T., Sato, F., Whiteoak, J.~B., et al.\ 1994, \apjl, 429, L77.

\bibitem[Haslam et al. 1982]{haslam+1982} Haslam, C.~G.~T., Salter, C.~J., et al.
1982, AAS, 47, 1   

\bibitem[Henshaw et al. 2022]{hen+2022} Henshaw, J. D., Barnes, A. T., Battersby, C., et al. 2022, arXiv:2203.11223, to appear in {\it Protostars and Planets} VII, ed. S. Inutsuka, et al., Univ. Arzona Press. 

\bibitem[Heywood et al. 2019]{hey+2019} Heywood I., et al., 2019, Natur, 573, 235 

\bibitem[Heywood et al. 2022]{hey+2022} Heywood, I., Rammala, I., Camilo, F., et al.\ 2022, ApJ, 925, 165. 



\bibitem[Kataoka et al. 2018]{ka+2018} Kataoka J., Sofue Y., Inoue Y., et al.
2018, Galax, 6, 27  

\bibitem[Kato et al. 2004]{kato+2004} Kato, Y., Mineshige, S., \& Shibata, K.\ 2004, \apj, 605, 307.

\bibitem[Kumar et al. 2006]{kumar+2006} Kumar, N., Kumar, P., \& Singh, S.\ 2006, \aap, 453, 1067.
 
  
\bibitem[LaRosa et al. 2004]{laro+2004} LaRosa, T.~N., Nord, M.~E., Lazio, T.~J.~W., et al.\ 2004, ApJ, 607, 302


\bibitem[Lang et al. 1999a]{la+1999a} Lang, C.~C., Anantharamaiah, K.~R., Kassim, N.~E., et al.\ 1999, ApJL, 521, L4 
\bibitem[Lang et al. 1999b]{la+1999b} Lang, C.~C.,  Morris M., Echevarria L., 1999, ApJ, 526, 727  
 
\bibitem[Morris et al. 1985]{mo+1985} Morris M., Yusef-Zadeh F., 1985, AJ, 90, 2511  

\bibitem[Nakashima et al. 2019]{nax+2019} Nakashima, S., Koyama, K., Wang, Q.~D., et al.\ 2019, ApJ, 875, 32 


\bibitem[Par{\'e} et al. 2019]{pa+2019} Par{\'e} D.~M., Lang C.~C., Morris M.~R., et al.
2019, ApJ, 884, 170 

\bibitem[Ponti et al. 2019]{po+2019} Ponti, G., Hofmann, F., Churazov, E., et al.\ 2019, Nat, 567, 347

\bibitem[Ponti et al. 2021]{po+2021} Ponti, G., Morris, M.~R., Churazov, E., et al.\ 2021, A\&Ap, 646, A66. 

\bibitem[Porter et al. 1994]{porter+1994} Porter, L.~J., Klimchuk, J.~A., \& Sturrock, P.~A.\ 1994, \apj, 435, 482.

\bibitem[Predehl et al. 2020]{pre+2020} Predehl, P., Sunyaev, R.~A., Becker, W., et al.\ 2020, Natur, 588, 227. 

\bibitem[Snowden et al. 1997]{sno+1997} Snowden S.~L., et al., 1997, ApJ, 485, 125 

\bibitem[Sofue 1977]{so1977} Sofue Y., 1977, A\&A, 60, 327 
\bibitem[Sofue 1978]{so1978} Sofue, Y.\ 1978, \aap, 67, 409
 
\bibitem[Sofue 1980]{so1980} Sofue Y., 1980, PASJ, 32, 79   
\bibitem[Sofue 1995]{so1995} Sofue, Y.\ 1995, PASJ, 47, 527  
 
\bibitem[Sofue 2000]{so2000}Sofue, Y. 2000, ApJ, 540, 224 
  

\bibitem[Sofue 2013]{so2013} Sofue, Y.\ 2013, \pasj, 65, 118. 
\bibitem[Sofue 2020a]{so2020th} Sofue, Y.\ 2020, \pasj, 72, L4.
\bibitem[Sofue 2020b]{so2020fe} Sofue, Y.\ 2020, \mnras, 498, 1335. 

\bibitem[Sofue 2022]{so2022} Sofue, Y.\ 2022, \mnras, 516, 907. 

\bibitem[Sofue et al. 1980]{so+1980} Sofue, Y., Furst, E., \& Hirth, W.\ 1980, \pasj, 32, 1 


\bibitem[Sofue et al. 1984]{so+1984} Sofue Y., Handa T., 1984, Natur, 310, 568  
\bibitem[Sofue et al. 2016]{so+2016} Sofue Y., Habe A., Kataoka J., et al.
2016, MNRAS, 459, 108    
\bibitem[Sofue et al. 2010]{so+2010} Sofue Y., Machida M., Kudoh T., 2010, PASJ, 62, 1191  

\bibitem[Sofue \& Reich 1979]{so+1979} Sofue, Y. \& Reich, W.\ 1979, \aaps, 38, 251
\bibitem[Sofue et al. 1987]{so+1987} Sofue, Y., Reich, W., Inoue, M., et al.\ 1987, \pasj, 39, 95 

\bibitem[Sormani \& Li 2020]{sormani+2020} Sormani, M.~C. \& Li, Z.\ 2020, \mnras, 494, 6030. 
 
\bibitem[Su et al. 2010]{su+2010} Su, M., Slatyer, T.~R., and Finkbeiner, D.~P.\ 2010, ApJ, 724, 1044      


\bibitem[Tsuboi et al. 1986]{ts+1986} Tsuboi, M., Inoue, M., Handa, T., et al.\ 1986, AJ, 92, 818 

\bibitem[Uchida 1970]{uch1970} Uchida Y., 1970, PASJ, 22, 341 
\bibitem[Uchida 1974]{uch1974} Uchida Y., 1974, SoPh, 39, 431  

 \bibitem[Xiao et al. 2008]{xiao+2008} Xiao, L., F{\"u}rst, E., Reich, W., et al.\ 2008, \aap, 482, 783. 
 
\bibitem[Vieu et al. 2020]{vieu+2020} Vieu, T., Gabici, S., \& Tatischeff, V.\ 2020, \mnras, 494, 3166.

\bibitem[Yusef-Zadeh et al. 1984]{yu+1984} Yusef-Zadeh F., Morris M., Chance D., 1984, Natur, 310, 557 
\bibitem[Yusef-Zadeh et al. 2019]{yu+2019} Yusef-Zadeh F., Wardle M., 2019, MNRAS, 490, L1 
\bibitem[Yusef-Zadeh et al. 2021]{yu+2021} Yusef-Zadeh, F., Wardle, M., Heinke, C., et al.\ 2021, \mnras, 500, 3142.

\bibitem[Yusef-Zadeh et al. 2022a]{yu+2022a} Yusef-Zadeh, F., Arendt, R.~G., Wardle, M., et al.\ 2022a, ApJ.L. 925, L18. 

 \bibitem[Yusef-Zadeh et al. 2022b]{yu+2022b} Yusef-Zadeh, F., Arendt, R.~G., Wardle, M., et al.\ 2022b, \mnras, 515, 3059.
\bibitem[Yusef-Zadeh et al. 2022c]{yu+2022c} Yusef-Zadeh, F., Arendt, R.~G., Wardle, M., et al.\ 2022c, arXiv:2208.11589  


\bibitem[Zhang et al. 2021]{zhang+2021} Zhang, M., Li, Z., \& Morris, M.~R.\ 2021, \apj, 913, 68. 

\end{thebibliography}
\end{document}